\documentclass[aip,reprint,onecolumn]{revtex4-1}

\usepackage[english]{babel} 
\usepackage{amsmath}
\usepackage{amssymb}
\usepackage{graphicx}
\usepackage{xcolor} 
\usepackage{amsbsy}
\usepackage{hyperref} 
\usepackage{xparse} 
\makeatletter\AtBeginDocument{\let\@elt\relax}\makeatother 

\usepackage{comment} 
\usepackage{ulem}

\newcommand{\ii}{\mathrm{i}}
\newcommand{\dd}{\,\mathrm{d}}
\newcommand{\w}{\omega}
\newcommand{\e}{\varepsilon}
\newcommand{\vp}{\varphi}
\def\pd2#1{\partial^2_{#1}}

\NewDocumentCommand{\evalat}{sO{\big}mm}{
  \IfBooleanTF{#1}
   {\mleft. #3 \mright|_{#4}}
   {#3#2|_{#4}}
}

\begin{document}

\title{Inferring oscillator's phase and amplitude response from a scalar signal exploiting test stimulation}
\author{Rok Cestnik}
\author{Erik T. K. Mau}
\author{Michael Rosenblum}
\affiliation{Department of Physics and Astronomy, University of Potsdam, 
Karl-Liebknecht-Str. 24/25, D-14476 Potsdam-Golm, Germany}
\date{\today}

\begin{abstract}
The phase sensitivity curve or phase response curve (PRC) quantifies the oscillator's reaction to stimulation at a specific phase and is a primary characteristic of a self-sustained oscillatory unit. Knowledge of this curve yields a phase dynamics description of the oscillator for arbitrary weak forcing. 
Similar, though much less studied characteristic, is the amplitude response that can be defined either using an {\it ad hoc} approach to amplitude estimation or via the isostable variables. Here, we discuss the problem of the phase and amplitude response inference from observations using test stimulation.
Although PRC determination for noise-free neuronal-like oscillators perturbed by narrow pulses is a well-known task, the general case remains a challenging problem.
Even more challenging is the inference of the amplitude response. This characteristic is crucial, e.g., for controlling the amplitude of the collective mode in a network of interacting units -- a task relevant to neuroscience.  
Here, we compare the performance of different techniques suitable for inferring the phase and amplitude response, particularly with application to macroscopic oscillators. We suggest improvements to these techniques, e.g., demonstrating how to obtain the PRC in case of stimuli of arbitrary shape. 
Our main result is a novel technique denoted by IPID-1, based on the direct reconstruction of the Winfree equation and the analogous first-order equation for isostable dynamics. The technique works for signals with or without well-pronounced marker events and pulses of arbitrary shape; in particular, we consider charge-balanced pulses typical in neuroscience applications. Moreover, this technique is superior for noisy and high-dimensional systems. Additionally, we describe an error measure that can be computed solely from data and complements any inference technique.
\end{abstract}

\maketitle

\section{Introduction}
Analysis and control of real-world oscillatory dynamics require inference of oscillator's parameters from observations. In particular, one can explore the system by applying a specifically designed perturbation and measuring the reaction. This paper discusses experiments extracting a self-sustained unit's phase and amplitude response. We concentrate on oscillators with weakly-stable limit cycles and stimulation with pulses of arbitrary shape.

In the first approximation in perturbation's strength, the oscillator's phase response is quantified by the phase sensitivity function or phase response curve (PRC) \cite{Winfree-80,Mackey_Glass-88,Rinzel-Ermentrout-98,Canavier-06}, a crucial characteristic of a self-sustained oscillator. Knowing the PRC one describes the oscillator's phase dynamics via the Winfree equation
\begin{equation}
  \dot\vp=\w+Z(\vp)p(t)\;,  
  \label{eq:Winfree}
\end{equation}
where $\vp$ is the phase, $\w$ is the natural frequency, $Z(\vp)$ is the PRC, and $p(t)$ is the external perturbation. 
Equation~(\ref{eq:Winfree}) predicts, e.g., the system's response to an arbitrary stimulus or synchronization of the oscillator by an external force. 
A description of oscillatory systems in terms of PRCs is useful in various fields, e.g., in computational neuroscience~\cite{Canavier-07}.

For known oscillator equations, one can obtain PRC by solving the adjoint problem~\cite{Ermentrout-book}. Estimation of PRC in an experiment is a less trivial task. The standard approach is to apply weak, narrow pulses at different phases and measure the phase shifts caused by stimulation~\cite{Mackey_Glass-88,Canavier-06}. Indeed, if the pulse at phase $\vp$ is Dirac's delta function, then the induced phase shift equals precisely $Z(\vp)$. It is easy to implement this approach for neuronal oscillators, where the distance between the spikes gives the variation of the period and hence, the phase shift. It is not that easy to exploit this technique to investigate oscillators without well-pronounced marker events that can be assigned a specific phase value. Furthermore, for any oscillator, the applied pulses should reasonably approximate the Dirac's delta pulses, which is often impossible because stimulation of living tissue should fulfill specific criteria, namely to be charge-balanced. 

A related, much more demanding problem is the inference of the amplitude response of an oscillator. This problem naturally arises, e.g., in model studies~\cite{Tass-99,*Tass-00,*Tass-01,*Tass_2001,*Tass-02, *Rosenblum-Pikovsky-04,*Rosenblum-Pikovsky-04a,
*Popovych-Hauptmann-Tass-05,
*Tukhlina-Rosenblum-Pikovsky-Kurths-07,*Hauptmann-Tass-09,*wilson2011,*Popovych-Tass-12,
*Lin_2013,*Zhou_2017,*Wilson-Moehlis-16,*Popovych_et_al-17,*Krylov-Dylov-Rosenblum-20,Montaseri_et_al-13,Holt_et_al-16,Rosenblum-20,Duchet_et_al-20} of the clinical technique known as deep brain stimulation (DBS)~\cite{Benabid_et_al-91,*Benabid_et_al-09,*Kuehn-Volkmann-17}. DBS aims to modulate the brain rhythm by stimuli delivered through implanted microelectrodes. Thus, a question arises: stimulation at which phase results in the most significant change of the amplitude \cite{Holt_Netoff-14,Holt_et_al-16,Duchet_et_al-20}. 

This paper critically assesses previously developed techniques and proposes two new approaches for estimating an oscillator's phase and amplitude response from observations, exploiting test pulses of arbitrary shapes. The first novel technique relies on a signal's instantaneous amplitude and phase and, hence,  is in the spirit of traditional time series analysis.  The second technique is model-based; it exploits recent advances in dynamical systems' description in terms of phases and isostable variables \cite{wilson_isostable_2016,wilson2018a,wilson2019} and reconstructs the equations for phase and, for the first time, for the isostable variable.
We exploit several test systems to probe our techniques' performance and compare them with those known in the literature. In particular, we concentrate on systems exhibiting signals without well-pronounced marker events, i.e., non-spiky signals, where the standard technique is inefficient.

The paper is organized as follows. In the rest of this section, we discuss the state-of-the-art and problem formulation in detail. In Section~\ref{sec:test} we present the tools for testing and comparing inference techniques. In Section~\ref{sec:methods_performance} we compare methods and their performance, including the novel reconstruction of the phase-isostable dynamics which is described in  Section~\ref{sec:phaseiso}. Section~\ref{sec:discus} summarizes and discusses the results. The appendix section introduces the phase-isostables representation and presents the details of the test systems and inference techniques.

\subsection{State of the art and problems to be solved}
Here, we specify the problems and introduce some notations.

\subsubsection{Estimating phase response in experiments}

The traditional approach relies on measuring the phase shift $\Delta\vp$ evoked by a pulse applied at phase $\vp$. 
However, in practice, the stimuli are not Dirac's delta pulses, so the inferred response function differs from $Z(\vp)$.  
Therefore, we consider general pulses ${\cal P}(t)$ of length $\delta$,
${\cal P}(t)=0$ for $t\notin [0,\delta]$.
In particular, we are interested in the charge-balanced stimuli that additionally fulfill the condition
\begin{equation}
\int\limits_{0}^{\delta} {\cal P}(t)\dd t=0\;.
\label{eq:chargebalanced}
\end{equation}
For neuroscience applications, such stimuli are required to avoid charge accumulation in living tissue. 
We denote the response to non-Dirac stimuli as empirical PRC  $Z_{\cal P}(\vp)$. Thus, the first problem is to relate
$Z(\vp)$ and $Z_{\cal P}(\vp)$ for a given ${\cal P}(t)$.
In the following, we infer the oscillator's response by perturbing it with the pulse train
\[  
p(t)=\sum_k {\cal P}(t-t_k)\;.
\]
We specify the choice of stimulation times $t_k$ in the test examples below. 
The phase shift caused by a stimulus also depends on the stimulus amplitude and hence, has to be normalized. For unipolar stimuli, it is natural to normalize by the integral $f=\int_{0}^{\delta} {\cal P}\dd t$, commonly referred to as the stimulus' "action"~\footnote{We highlight that in the context of electrical stimulation, e.g., in neuroscience applications, $p(t)$ is a current or a voltage and the introduced action has the physical meaning of electrical charge.}. 
For bipolar charge-balanced pulses the normalization is ambiguous; we choose $f=\frac{1}{2}\int_0^\delta |{\cal P}|\dd t$. 
Thus, $Z_{\cal P}=\Delta\vp/f$.

The success of the standard technique is due to the presence of well-defined spikes, typical for neuronal oscillators. Consider now a system that exhibits nearly sinusoidal oscillation. Although no well-defined markers exist, threshold-crossing events can also determine the points of a constant phase value~\cite{PhysRevE.65.051110}. However, the error in phase estimation should be higher than in the case of neuron-like oscillators, especially in the presence of noise. 
Therefore, the natural idea is to avoid determining the marker events and estimate the instantaneous phase. The first step in this direction has been done by Holt et al.~\cite{Holt_Netoff-14,Holt_et_al-16}. They suggested fitting sine waves to the several oscillation periods before and after each stimulus. The phase shift caused by the stimulus is then easily obtained from these two sines.
Duchet et al.~\cite{Duchet_et_al-20} approached this problem using the Hilbert Transform: they estimated the instantaneous phase before and after the stimulus and, in this way, computed the PRC.~\footnote{Instantaneous phase and amplitude of a real-valued signal $s(t)$ are the argument and absolute value
of the complex-valued analytical signal $s(t)+\ii s_H(t)$, where $s_H$ is the Hilbert Transform of $s$, see, e.g.,~\cite{pikovsky2001,King-09,Feldman-11}.
}
Unfortunately, neither of the techniques published in~\cite{Holt_Netoff-14,Holt_et_al-16,Duchet_et_al-20} has been tested on an oscillatory model with known PRC; therefore, the performance of these techniques is unclear. Below, we test these techniques and propose a different, more precise approach.

\subsubsection{Estimating amplitude response in experiments}

The problem of amplitude response estimation appears, e.g., in the context of DBS, where the goal is to modulate the brain rhythm by weak pulses. Such modulation is possible only if the amplitude's perturbation decays slowly so that the effects of several pulses accumulate. For a dynamical model, it means that the oscillator's limit cycle is weakly stable, i.e., its largest Floquet multiplier is close to one (Floquet exponent close to 0).  
A popular model of brain rhythm generation is a large neuronal population exhibiting collective mode oscillation due to the synchronization of the population's elements. If the amplitude of the collective mode is not too large, i.e., the system is close to the synchronization transition point, then the limit cycle corresponding to the collective oscillation is weakly stable even if the cycles of individual neurons are strongly stable. This consideration motivates our analysis of the amplitude response using the models with a weakly stable cycle. 

The first problem with the amplitude response is its definition. While the phase of a limit-cycle oscillator is unambiguously (up to an additive constant) defined via isochrons, there is no universal definition for the amplitude variable. A pragmatic approach~\cite{Duchet_et_al-20} is to estimate the amplitude of an observed signal using the Hilbert Transform. Comparing the amplitude before and after the stimulus for different phases of stimulation, Duchet et al.~\cite{Duchet_et_al-20} in this way introduced the amplitude response curve, ARC. 

Another approach to characterizing the amplitude variations exploits the notion of isostables~\cite{wilson2016, wilson2018, wilson2019}. In this framework, the Winfree Eq.~(\ref{eq:Winfree}) is complemented by an equation for the isostable variable $\psi$ that can be interpreted as a deviation from the limit cycle:
\begin{equation}
    \dot\psi=\kappa\psi +I(\vp)p(t)\;.
    \label{eq:iso}
\end{equation}
The real-valued parameter $\kappa$ (Floquet exponent) determines the stability of the cycle and is negative for stable cycles. The function $I(\vp)$ quantifies the phase dependence of the stimulation's effect. We denote $I(\vp)$ as the isostable response curve, IRC. For an introduction to the phase-isostable representation, see Appendix~\ref{app:phipsi}.
We remind that Eqs.~(\ref{eq:Winfree},\ref{eq:iso}) represent the first-order approximation in the perturbation's strength. 

\bigskip
This paper elaborates on the phase and amplitude response inference employing test stimulation. 
To evaluate the performance of different techniques, we test them on systems for which this response is known. First, we analyze the existing signal analysis techniques and propose some improvements.
Next, we present the method for inference of the phase-isostable Eqs.~(\ref{eq:Winfree},\ref{eq:iso}) for the first time.

\section{Tools for testing and comparing inference techniques}
\label{sec:test}

In this section, we first suggest a simple approach for relating the response curve to stimuli of arbitrary shape $Z_{\cal P}(\vp)$ to the infinitesimal PRC $Z(\vp)$. Then we design and present the test systems. And finally, we introduce the error measures we will be using.

\subsection{Relation between empirical $Z_\mathcal{P}$ and infinitesimal $Z$ PRCs}\label{sec:deconvolution}
From the infinitesimal PRC $Z(\vp)$, one can compute the empirical PRC $Z_\mathcal{P}(\vp)$ as a response to an arbitrary stimulus $\mathcal{P}$. One simply has to evaluate the effective phase shift for the chosen stimulation $\mathcal{P}$ at every phase by integrating the Winfree phase equation~\eqref{eq:Winfree}. For example, let us focus on evaluating the phase response for stimuli $\mathcal{P}$ at a particular phase $\vp^*$. 
One first needs to compute the phase in time for the duration of the pulse by solving the Winfree equation~\eqref{eq:Winfree} using initial condition $\vp(t=0) = \vp^*$. 
Then the empirical PRC at the chosen phase corresponds to the phase shift that is induced by $\mathcal{P}$ over the duration of the stimulation:
\begin{equation}
Z_\mathcal{P}(\vp^*) = \frac{1}{f} \int\limits_{0}^\delta Z(\vp(t))\mathcal{P}(t) \dd t \;,
\end{equation}
where $f$ is the action of the pulse.

The inverse problem of how to determine the infinitesimal response $Z(\vp)$ when given the empirical one $Z_\mathcal{P}(\vp)$ and stimulation shape $\mathcal{P}$, is in general, much harder. However, for pulses with small action $f$, the empirical response $Z_\mathcal{P}(\vp)$ is well approximated as the convolution of the infinitesimal curve $Z(\vp)$ and the pulse shape $\mathcal{P}(t)$. In Fourier space, convolution corresponds to multiplication, meaning the inverse operation is represented by division. One can, therefore, easily deconvolve an empirical response to the infinitesimal one by dividing the Fourier representations of $Z_{\mathcal{P}}$ and $\mathcal{P}$; we present the technique in more detail in Appendix~\ref{app:inverse}.
There we also present an algorithm that does not assume the smallness of $f$.
The transformation $Z_{\cal P}\to Z$ is generic and can be combined with any technique that infers an empirical response.

\subsection{Test models}
\subsubsection{Two-dimensional test models}

Our first test model is the Stuart-Landau (SL) oscillator. The reasons for using this system are threefold. First, the parameters of this system explicitly govern the limit cycle's stability and shape of isochrons. Second, this system's phase and amplitude response can be obtained analytically. Third, this system represents the normal form of the Hopf bifurcation and, therefore, serves as the model equation for the collective mode of an oscillator population close to the synchronization transition point.
The system's equations are:
\begin{equation}
\begin{aligned}
   \dot{x} &= \mu x - \eta y - (x^2 + y^2)(x-\alpha y) +\sigma\xi_x(t)+\cos\beta\cdot p(t)\;,\\
   \dot{y} &= \mu y + \eta x - (x^2 + y^2)(y+\alpha x) +\sigma\xi_y(t)+\sin\beta \cdot p(t)
   \label{eq:SL_cartesian} \;,
\end{aligned}
\end{equation}
where the parameter $\mu$ governs the limit cycle's stability as it is directly related to its Floquet exponent by $\kappa = -2\mu$ and $\alpha$ is the non-isochronicity parameter. The frequency of the limit-cycle oscillation is given by $\omega := \eta - \alpha \mu$. $\sigma$ is the noise strength, and $\xi_{x,y}$ are two realizations of the Gaussian white noise with zero mean and unit variance. $p(t)$ is the external perturbation specified in the test examples below, and parameter $\beta$ determines the perturbation's direction with respect to $x$ and $y$ variables.
The solution of the noise-free and unperturbed SL system is a sine with the amplitude $\sqrt{\mu}$.
For definiteness, throughout this paper, we assume that the perturbation is state-independent in the chosen Cartesian coordinates~\eqref{eq:SL_cartesian}.
Then, the SL system~(\ref{eq:SL_cartesian}) has PRC (see, e.g., \cite{Rosenblum-Pikovsky-19a})
\begin{equation}
    Z(\vp) = -\frac{1}{\sqrt{\mu}} \left( \sin(\vp-\beta) + \alpha \cos(\vp-\beta)\right)\;
    \label{eq:SL_PRC}
\end{equation}
and IRC
\begin{equation}
    I(\vp) = \frac{2}{\sqrt{\mu}}\cos(\vp-\beta) \;.
    \label{eq:SL_IRC}
\end{equation}
For the SL system, we also define the empirical amplitude response $A(\vp)=R_e/R_s$, where $R_{s,e}$ are the values of the amplitude variable $R$ immediately before and after the applied pulse. For a narrow unipolar pulse, this expression reads:
\begin{equation}
A(\vp)= 1 + \frac{f \cos(\vp-\beta)}{\sqrt{\mu}}\;,
\label{eq:SL_ARC}
\end{equation}
where $f$ is the pulse's action. 
For the derivation of IRC and ARC for the SL model, and their 
interrelation, see Appendix~\ref{app:phipsiSL}. 

\begin{figure}
    \includegraphics[width=0.95\textwidth]{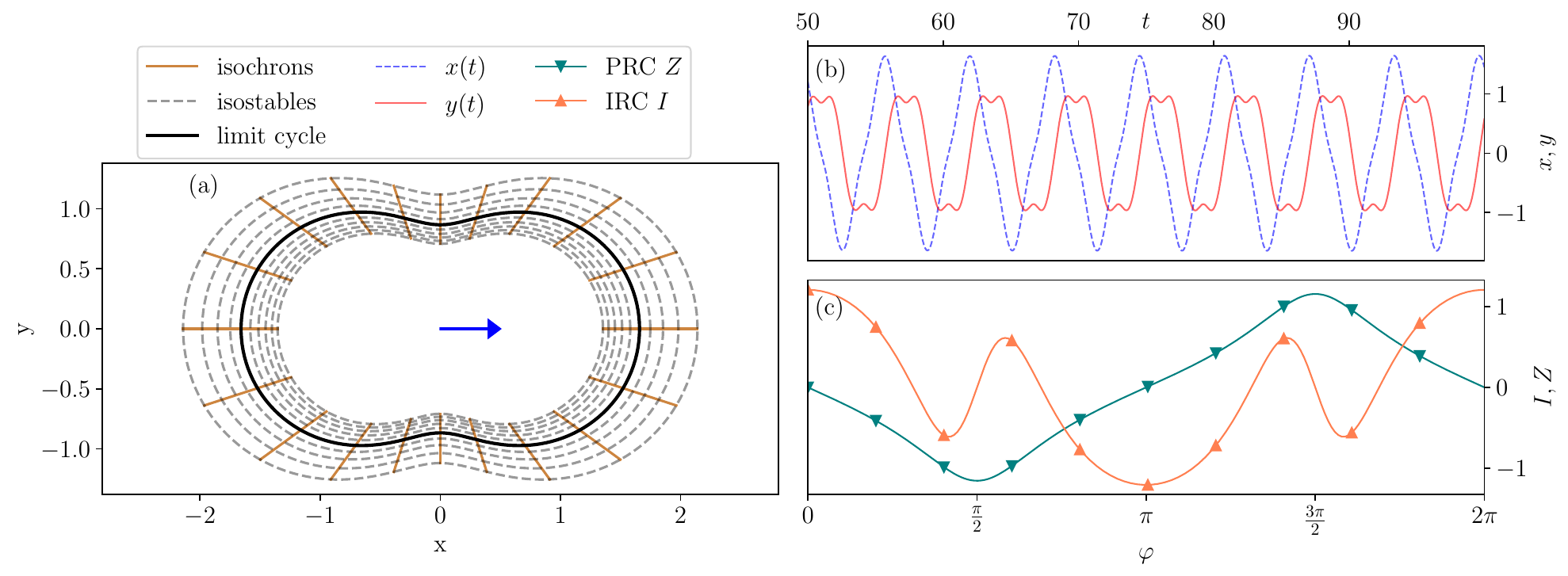}
    \caption{ Panel (a) shows the phase space of the modified SL model (\ref{eq:mSL_IRC}) with the stable limit cycle as the solid black line, the isochrons as solid brown lines, and the isostables as dashed grey lines. The blue arrow indicates the direction in which stimulation enters the system, i.e., in $x$-direction as $\beta=0$. An epoch of the time evolution of the $x$- and $y$-component on the limit cycle is illustrated in panel (b) as dashed and solid lines, respectively. In panel (c), the PRC $Z$ (greenish, triangles down) and IRC $I$ (orange, triangles up) for forcing in $x$-direction are shown. Parameter values are $\w=1$, $\kappa=-0.1$, $\alpha=0.0$, $r=0.75$. }
    \label{fig:infinity_oscillator}
\end{figure}

We modify the SL oscillator to obtain the second test model that provides a non-harmonic solution but is still analytically tractable. We denote this system as the modified SL oscillator (mSL). The system's equations that we derive in Appendix~\ref{app:phipsiEM} include the frequency of the limit cycle's oscillation $\w$, the Floquet exponent $\kappa$, and non-isochronicity $\alpha$ as parameters. With perturbations, these equations read: 
\begin{equation}
\begin{aligned}
    \dot{x} &= 
    \w \left[xC(x,y)-y \right]
    + \frac{\kappa}{2}\left[D(x,y)-1\right]\left[x + \alpha \left( xC(x,y)-y \right) \right] +\sigma\xi_x(t) + \cos \beta \cdot p(t)\;,\\
    \dot{y} &=  
    \w \left[ yC(x,y)+x \right]
    + \frac{\kappa}{2}\left[D(x,y)-1\right]\left[y + \alpha \left( yC(x,y)+x \right) \right] +\sigma\xi_y(t) + \sin \beta \cdot p(t)\;,
\end{aligned}
\label{eq:infosc}
\end{equation}
where the functions $C$ and $D$ are:
\begin{align}
    C(x,y) &= - \frac{2xy}{(r+2)x^2 + ry^2} \;,\\
    D(x,y) &= \frac{(x^2 + y^2)^2}{(r+2)x^2 + ry^2}\;,
\end{align}
and $r$ is a positive parameter determining the shape of the limit cycle in state space.
System~(\ref{eq:infosc}) 
has IRC 
\begin{align}
    I(\vp) = 
    2 \frac{(r+1)\cos(\vp-\beta) + \cos(3\vp-\beta)}{(r+2\cos^2(\vp))^\frac{3}{2}}
    \label{eq:mSL_IRC}
\end{align}
and PRC  
\begin{align}
    Z(\vp) = -\frac{\sin(\vp-\beta)}{\sqrt{r+2\cos^2(\vp)}} - \alpha \frac{(r+1)\cos(\vp-\beta) + \cos(3\vp-\beta)}{(r+2\cos^2(\vp))^\frac{3}{2}}\;.
    \label{eq:mSL_PRC}
\end{align}
The system's dynamics are illustrated in Fig.~\ref{fig:infinity_oscillator}.

\subsubsection{High-dimensional test model}
\begin{figure}
    \includegraphics[width=0.8\textwidth]{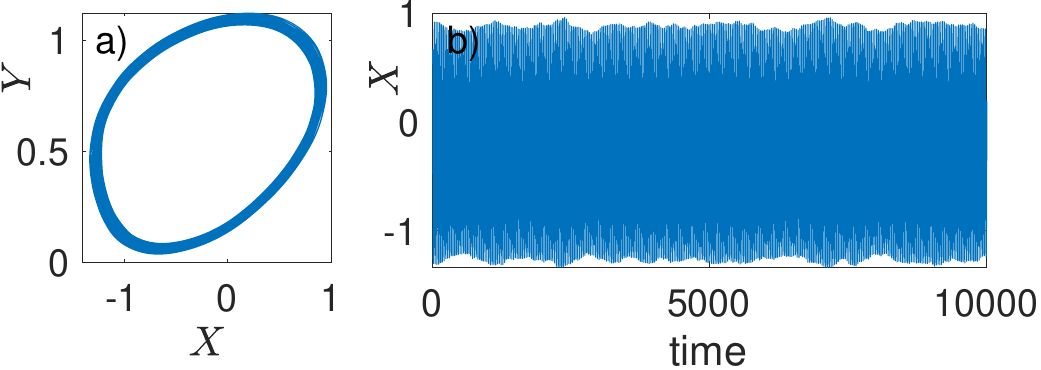}
    \caption{Collective  oscillation on an ensemble of globally coupled units, Eqs.~(\ref{eq:bvdp}). (a) The trajectory in mean-field coordinates (see text). (b) Time dependence of the mean-field oscillation exhibits amplitude modulation, typical, e.g., for band-pass filtered brain activity, cf. Fig. 1 in Ref.~\cite{Duchet_et_al-20}.  }
    \label{fig:ensemb_traj}
\end{figure}

Our third test system is a simple model of macroscopic neuronal oscillation. We take a system of $N$ globally coupled Bonhoeffer--van der Pol oscillators:
\begin{equation}
\begin{aligned}
\dot{x}_k &= x_k-x_k^3/3 - y_k +J_k +\e X +  p(t)\;,\\
\dot{y}_k &= 0.1 (x_k-0.8y_k+0.7)\;,
\end{aligned}
\label{eq:bvdp}
\end{equation}
where $k$ is the oscillator index, $k=1,\ldots,N$, and the term $\e X$ describes the mean-field coupling, where $X=N^{-1}\sum_k x_k$. In the following, we take $N=1000$.
The oscillators' frequencies are determined by the parameter $J_k$ 
that is Gaussian-distributed with mean $0.6$ and standard deviation $0.1$. The coefficient $\e$ explicitly describes 
the interaction between the ensemble elements. 
We choose $\e=0.023$; for this parameter's value, the system exhibits collective chaos so that the mean-field oscillation is amplitude-modulated. Figure~\ref{fig:ensemb_traj} presents the two-dimensional trajectory in coordinates $X$ and $Y=N^{-1}\sum_k y_k$ in (a) and $X(t)$ in (b).
With this test example, we imitate the natural variability of real-world signals. 

\subsection{Error measures}
We define two error measures with which we quantify the goodness of inference. If the true response curve is known from the theory, we evaluate the goodness of the inference by computing the normalized $L^2$ distance between the inferred and true curves. We denote these errors with $L_{Z,A,I}$, where the subscript indicates which curves we are comparing (either PRC, ARC, or IRC). Thus, for error of, e.g., the PRC recovery, we obtain
\begin{equation}
    L_Z 
    = \sqrt{ \frac{\langle (Z - Z^\text{rec})^2\rangle }{\langle (Z-\langle Z \rangle)^2\rangle} }
    \;,
    \label{eq:errorL}
\end{equation}
where $\langle \cdot \rangle = (2\pi)^{-1}\int\limits_{0}^{2\pi} (\cdot) \dd \vp$ denotes the average value of a function, and $Z^\text{rec}$ is the inferred (recovered) PRC.
The error value is $L_{Z,A,I}=0$ if the two curves coincide, it is of order 1 if the two curves are of the same order of magnitude but not similar and can also take higher values if one curve is significantly larger on average. 

The other measure quantifies how well the inferred curve represents the dynamical model. Since the dynamics of phase and amplitude are distinct (see Eqs.~\eqref{eq:Winfree} and \eqref{eq:iso}, respectively) the definition of the corresponding error measures differs as well. 
We first introduce the error measure for the phase response and later in Sec~\ref{sec:IRC_inf} when introducing the method we also specify an analog for the isostable variable error. 
Note that the inferred phase response curve $Z(\vp)$ in conjunction with the natural frequency $\omega$ can be used in Eq.~\eqref{eq:Winfree} to reproduce $\vp(t)$ for a given realization of stimulation $p(t)$ by means of numerical integration. 
In the first step, we threshold the observed signal to determine time events $\tau_i$ that correspond to the same phase; for details, see Appendix~\ref{app:threshold_crossing}. Without loss of generality, we set this phase to zero. Next, starting at $\tau_i$ we reproduce the phase evolution up to the time $\tau_{i+1}$.
In the ideal noise-free case, if the stimulation is weak and the inferred curve is exact, the reproduced phase $\vp(\tau_{i+1})=\Phi_i$ equals $2\pi$
(or, equivalently, zero since we consider the wrapped phase). 
In practice, this reproduced phase $\Phi_i$ deviates from $2\pi$, 
and it is precisely this deviation that we use to quantify the inference's quality. We define the error of PRC reconstruction 
$E_{Z}$ as the standard deviation of $\Phi_i$ from $2\pi$:
\begin{equation}
    E_Z = \langle (\Phi_i-2\pi)^2 \rangle^{1/2}\;,
    \label{eq:Z_error}
\end{equation}
where $\langle\cdot\rangle$ denotes averaging over the index $i$. 
The value of $E_{Z}$ should be compared to a measure of the signal's irregularity.  We evaluate the latter by a measure that is proportional to the standard deviation of the inter-event intervals $T_i=\tau_{i+1}-\tau_i$
(instantaneous periods):
\begin{equation}
E_{Z0}=\frac{2\pi}{\langle T_i\rangle} 
\langle (T_i-\langle T_i\rangle)^2\rangle^{1/2}\;.
\label{eq:Z0_error}
\end{equation}
A natural measure of the inference's quality is the ratio  ${E_Z}/{E_{Z0}}$ since it roughly describes how much of the signal's irregularity is explained by the inferred phase model. If one considers a PRC that is identically zero, this ratio equals one.

We emphasize an essential advantage of the error measure ${E_Z}/{E_{Z0}}$. The computation of the 
$L^2$-based measure Eq.~(\ref{eq:errorL}) requires knowledge of the ground truth and, therefore, helps only in testing the techniques on model data from limit-cycle oscillators. In contrast, the error measure ${E_Z}/{E_{Z0}}$ is obtained solely from our inferred model and the data itself. This makes it a helpful tool for experimental data analysis. 

\section{Methods and their performance}\label{sec:methods_performance}
We formulate the inference problem as follows. Suppose we perturb the system by applying some known stimulation $p(t)$
and measure the system's scalar output $s(t)$. We process $s(t)$ to infer the PRC $Z(\vp)$ and IRC $I(\vp)$ or ARC $A(\vp)$ as  specified below. Generally, we are free to construct $p(t)$, e.g., as a sequence of pulses. However, some constraints exist in specific settings, e.g., the pulses must be charge-balanced in neuroscience applications. 
In this section, we first review the standard PRC inference technique, followed by testing the performance of other methods in use. Here we also describe and test our approach to the problem. 

\subsection{Standard approach (phase response only)}\label{sec:standard_technique}

The standard technique is very efficient for neuronal oscillators exhibiting slow and fast motion. A spike corresponds to an epoch of fast motion, where the isochron density in the phase space is low. Therefore, since spike detection via threshold-crossing is weakly dependent on the threshold, the precision of the phase estimation is high. Consider now a system without slow-fast motion that exhibits nearly sinusoidal oscillation. 
To trace the variation of the period, we determine states with equal phases employing threshold-crossing~\footnote{Alternatively, one can detect the signal's maxima; this approach corresponds to threshold-crossing for the derivative.}, cf.~\cite{PhysRevE.65.051110}. 
For example, we apply a stimulus and look for the events with zero crossings from below. However, if the limit cycle is not strongly stable, observing such an event immediately after the stimulus's application at $t_s$ does not suffice. Indeed, we have to wait until the system returns to the limit cycle. Thus, we take the instant $\tau_n$ of the $n^\text{th}$ threshold-crossing event and compute 
\begin{equation}
  Z_{\cal P}(\vp)= \frac{2\pi}{f}\frac{nT-(\tau_n-\tau_0)}{T}\;,
\label{eq:standard}  
\end{equation}
where $\tau_0$ is the threshold-crossing event preceding the stimulus, and the phase of the stimulus application is $\vp=2\pi(t_s-\tau_0)/T$ where $T$ is the natural period. The choice of $n$ depends on the stability of the limit cycle, quantified with the Floquet exponent $\kappa$; as a rule of thumb, we suggest $n \kappa T \gg 1$. However, in practice, $\kappa$ is unknown, and $n$ shall be chosen by trial and error. To obtain the function $Z_{\cal P}(\vp)$, we repeat the perturbation $\mathcal{P}$ for different $\vp$. To this end, we either choose random intervals between stimuli or stimulate periodically with the period incommensurate with $T$. In any case, the stimulation interval shall be large enough to ensure observation of $n$ threshold-crossing events after each applied pulse. 

\begin{figure}
    \centering
    \includegraphics[width=0.36\paperwidth]{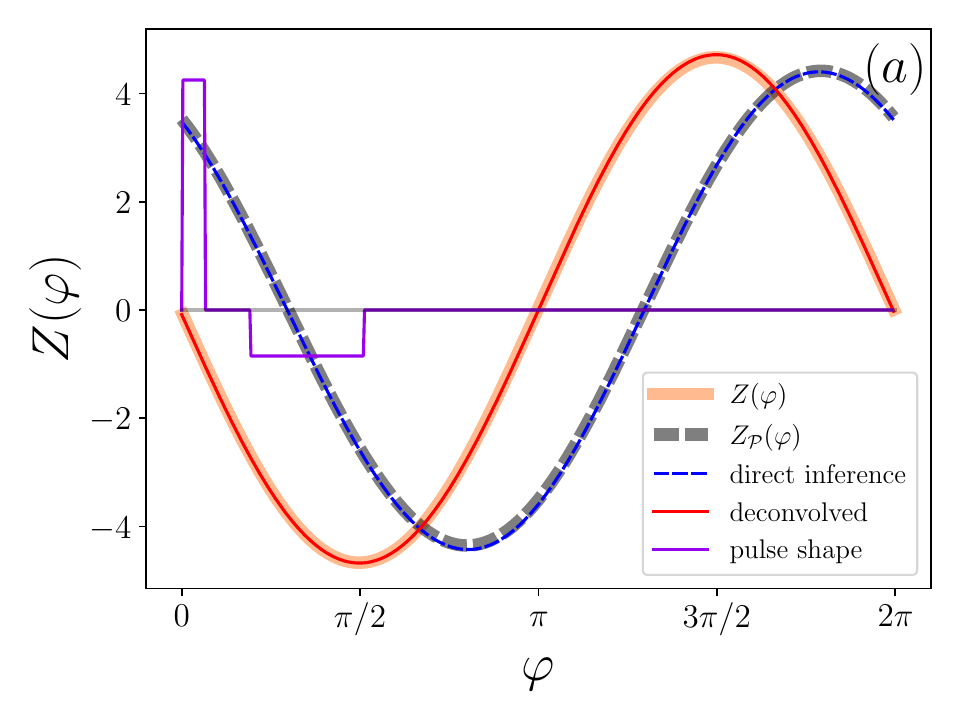}
    \includegraphics[width=0.36\paperwidth]{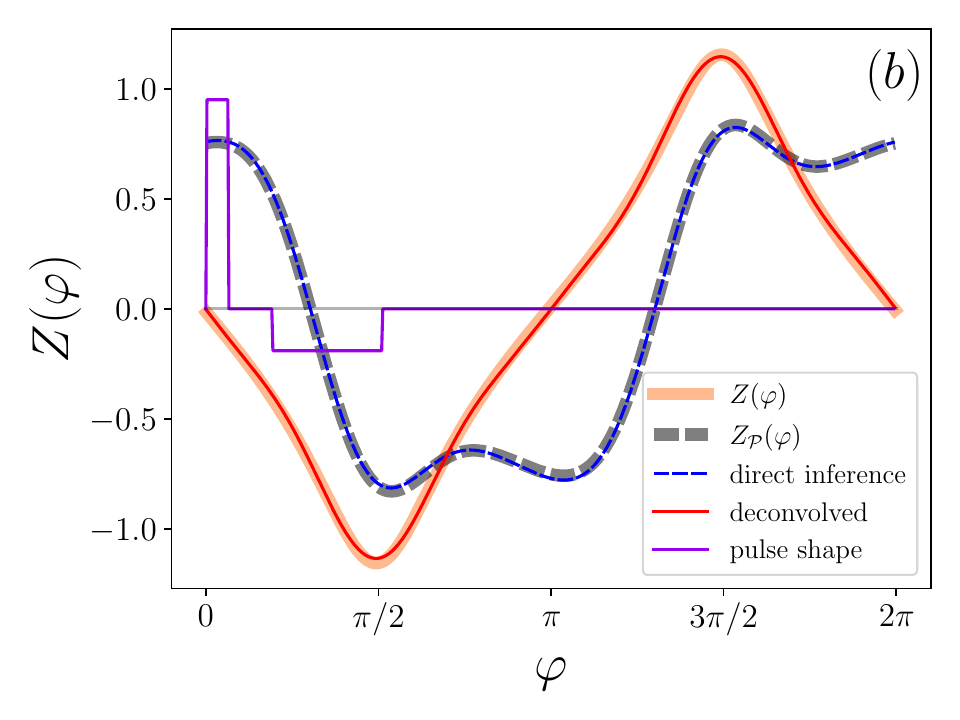}\\
    \caption{Inference of the phase response by the standard technique, for charge-balanced stimulation and for two test oscillators, (a) Stuart-Landau and (b) its modification, ~Eq.~\eqref{eq:infosc}. Orange (full) and gray (dashed) bold transparent curves depict the theoretical curves $Z(\vp)$ and $Z_{\cal P}(\vp)$.  Blue (dashed) and red (full) thin curves show the results of the direct inference and of the deconvolution (see text for explanations).
    The purple line demonstrates the  pulse shape as a function of $t=\frac{\vp}{2\pi}T$ (as it would appear next to an unperturbed signal). The pulse is vertically scaled to fit the plot. 
    We see a very good correspondence between the theoretical and inferred curves.}
    \label{fig:prc_standard}
\end{figure}

We illustrate the inference via the standard technique in Fig.~\ref{fig:prc_standard} where we consider a charge-balanced stimulation and infer the response for the two test models presented earlier, see Eqs.~(\ref{eq:SL_cartesian},\ref{eq:infosc}). 
For the SL system~\eqref{eq:SL_cartesian} the parameter values are: $\omega = 1$, $\kappa = -0.1$, $\alpha = -0.3$ and for the mSL system~\eqref{eq:infosc} they are $\omega = 1$, $\kappa = -0.1$, $\alpha=0$, $r =0.75$. There is no noise, $\sigma = 0$. For the observable, we take the $x$ variable, and $n=3$ threshold-crossing events after stimulation were considered for the phase shift determination. The charge-balanced pulse used has a short interval of positive stimulation (duration 0.2), followed by an interval without stimulation (duration 0.4), and then a longer interval of negative stimulation (duration 1.0). The amplitude of stimulation depends on the oscillator, the action $f = \frac{1}{2} \int |\mathcal{P}| \dd t$ is 0.01 for SL and 0.07 for mSL, see Appendix section~\ref{app:forcing_gen} for details on perturbation generation. We show the following curves: (i) theoretical PRC $Z(\vp)$ according to Eqs.~(\ref{eq:SL_PRC}) and (\ref{eq:mSL_PRC}), 
(ii) theoretical response to charge-balanced pulses 
$Z_{\cal P}$ - the ground truth, 
(iii) inferred empirical PRC $Z_{\cal P}$
obtained via Eq.~(\ref{eq:standard}) (``direct inference'')~\footnote{The empirical curve is an $8^\text{th}$-order Fourier fit of experimental points.}, 
and (iv) the result of the deconvolution of the empirical PRC $Z_{\cal P}$ as described in Appendix~\ref{app:inverse} (``deconvolved''). For this simple case, we see a nearly perfect reconstruction of the effective PRC $Z_{\cal P}$ (compare blue and gray curves in Fig.~\ref{fig:prc_standard}) as well as a successful deconvolution to the infinitesimal curve (compare orange and red). Since we used bipolar pulses, the empirical PRC resembles the negative derivative of the infinitesimal curve (compare gray with orange).

We emphasize that there is no analog to the standard technique for inferring the amplitude response. 

\subsection{Inferring response measuring instantaneous phase and amplitude}
\label{sec:signal}

The idea of the approach is straightforward. Suppose we perturb the oscillator by a sequence of pulses at instants $t_k$ and measure the system's output $s(t)$. Computing the instantaneous phase and amplitude of $s(t)$ before and after each pulse, we obtain the phase shift and amplitude variation as functions of the stimulation phase. To obtain a reasonable estimation of these functions, we have to apply the pulses at different phases -- we ensure this by either using stimulation with a period incommensurate with that of the unperturbed oscillator or by choosing random $t_k$.

Let the instantaneous phase and amplitude immediately before and after a finite-width stimulus be $\vp_{s,e}$ and $a_{s,e}$, respectively, where indices $s$ and $e$ stand for ``start'' and ``end''.  From these quantities, we obtain the empirical PRC $Z_{\cal P}(\vp_s)=(\vp_e-\vp_s-\w \delta)/f$, where $\delta$ is the pulse's width, $\w$ is the frequency of the unperturbed oscillation, and $f$ is the normalization factor. Next, following Duchet et al.~\cite{Duchet_et_al-20}, we introduce the empirical amplitude response curve (ARC) as $A(\vp)=a_e/a_s$, to be distinguished from IRC. Below, we explore the performance of three different techniques.

\subsubsection{Fitting the signal by a sine (phase response only)}

\begin{figure}
    \centerline{\includegraphics[width=0.4\paperwidth]{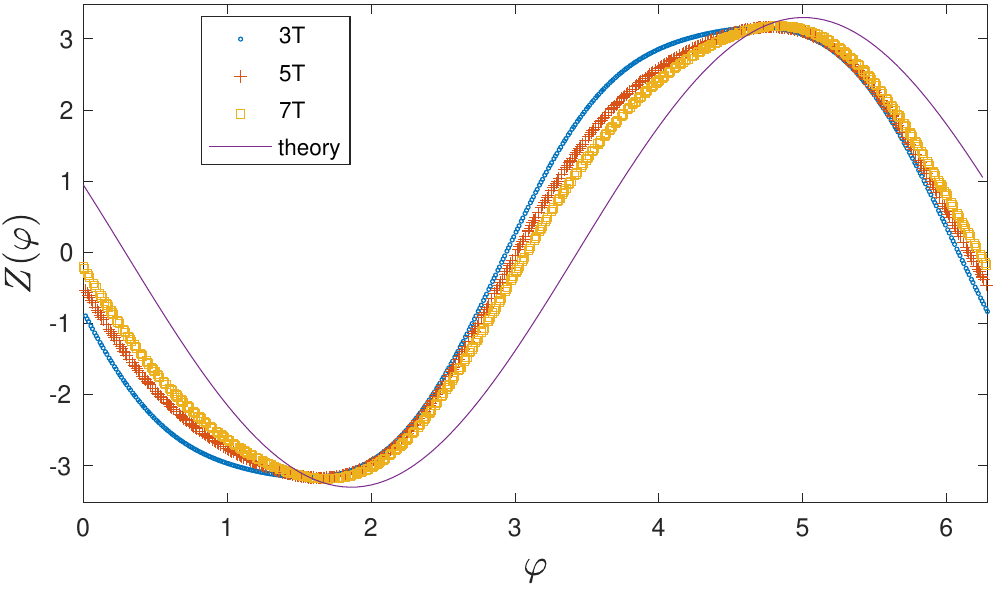}}
\caption{PRC obtained via fitting a cosine prior and after the stimulus for different lengths of the fitting interval (the interval is given in periods of the unperturbed system).  The solid line depicts the theoretical PRC $Z(\vp)$, see Eq.~\eqref{eq:SL_PRC}.
}
    \label{fig:PRC_Ffit}
\end{figure}

Holt et al.~\cite{Holt_Netoff-14,Holt_et_al-16} suggested fitting a harmonic to several periods of the signal $s(t)$ before the stimulus and another harmonic to an interval of the same length $T_{\text{fit}}$ after the stimulus. Then, they exploited the Fourier Transform of both fitted functions to find the phase shift evoked by the stimulus. Obviously, this technique does not provide information on the amplitude response. Indeed, the perturbation in the amplitude normally decays rather quickly and cannot be captured by a fit over several oscillation periods. We illustrate the performance of this technique on the test model (\ref{eq:SL_cartesian}) in Fig.~\ref{fig:PRC_Ffit}, computing the phase shift in the time domain. Suppose the pulse of length $\delta$ occurs
at $t_s$. Let the fitted functions be $s(t)\approx\ a_1\cos[\w (t-(t_s-T_\text{fit})) +\chi_1]$ for $t_s-T_\text{fit}<t<t_{s}$ and $s(t)\approx\ a_2\cos[\w (t-(t_s-T_\text{fit})) +\chi_2]$ for $t_s+\delta<t<t_s+\delta+T_\text{fit}$. 
The phase of the second cosine at $t=t_s+\delta$ is $\vp=\w(T_\text{fit}+\delta)+\chi_2$. If there were no stimulus, the phase of the first cosine at this point would be $\vp=\w(\delta+T_\text{fit})+\chi_1$. Hence, the phase shift is $\Delta\vp=\chi_2-\chi_1$.~\footnote{Note that the oscillation frequency $\w$ shall be either determined by fit or set to be equal to the central frequency of the bandpass filter, see~\cite{Holt_Netoff-14,Holt_et_al-16}.} 

Naturally, this technique applies to the signals close to  harmonic oscillation. Therefore, we test it on the SL oscillator only. We use the same parameters as used to illustrate the standard technique, see Fig.~\ref{fig:prc_standard}. For simplicity, we use unipolar rectangular pulses with amplitude $0.1$ and width $\delta=0.03$, therefore having action $f=0.003$. Such a stimulus is a good approximation for the delta-pulse, hence, we can expect a good correspondence of the inferred PRC with the theoretical one. However, the results summarized in Fig.~\ref{fig:PRC_Ffit} demonstrate a limited precision of the approach.
As another drawback of the technique, we mention that, since fitting requires several periods before and after the stimulus, the stimulation pulses shall be relatively rare, and hence, the time interval required for PRC estimation  is rather long.

\subsubsection{Estimating phase and amplitude using the Hilbert Transform}

A typical way to obtain the narrow-band signal's instantaneous phase and amplitude is to use the Hilbert Transform (HT). Duchet et al.~\cite{Duchet_et_al-20} exploited HT to compute the phase and amplitude response of tremor oscillation with DBS of the thalamus. They took the instantaneous phase and amplitude immediately before and after the stimulus for this computation~\footnote{In their experiments,
Duchet et al.~\cite{Duchet_et_al-20} used a complex stimulus consisting of 25 bursts of high-frequency pulses of a total length of 5 s.}.

\begin{figure}
\centerline{\includegraphics[width=0.6\paperwidth]{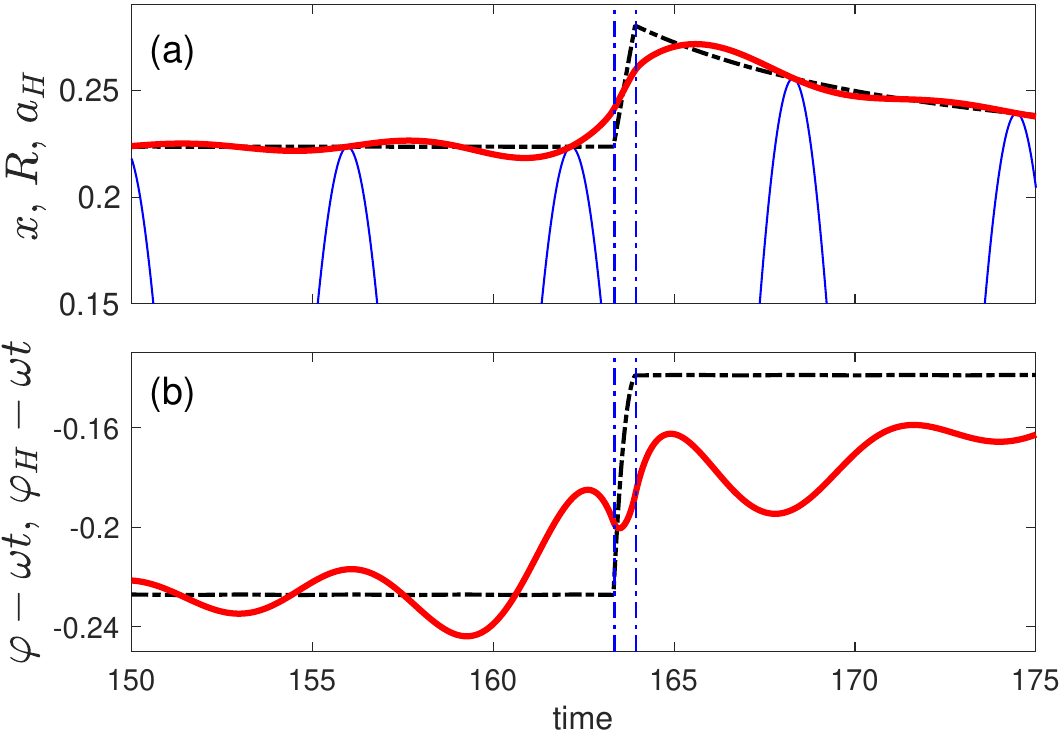}}
\caption{HT-based amplitude and phase of the perturbed SL oscillator. 
(a) Black dashed-dotted curve shows the time dependence of the amplitude variable
$R=\sqrt{x^2+y^2}$ of the SL system, and the bold red curve shows the instantaneous amplitude $a_H$ computed via the Hilbert Transform. The thin blue curve depicts the observation $x(t)$.
(b) Here, the black dashed-dotted curve shows the evolution of the oscillator's phase $\vp$, while the red curve illustrates the HT-based phase $\vp_H$ (for both phases, we subtract the average growth with the frequency $\w$).
The vertical lines in all panels indicate the beginning and end of the rectangular pulse. In the absence of stimulation, the HT amplitude $a_H(t)$ reproduces $R(t)$, but the discrepancy is essential before and after the pulse. For the phases, the effect is even more pronounced.
}
    \label{fig:HT1}
\end{figure}

However, being a nonlocal operation, the HT is not suitable for measuring response to a pulse, e.g., the HT shows the systems' reaction even {\it before} the stimulus begins.
We illustrate this property of the HT by perturbing the SL oscillator with a rectangular pulse. For a better visibility, we choose a wide pulse (amplitude 0.1, $\delta=0.6$). 
The parameters of the SL system are the same as before. Figure~\ref{fig:HT1} illustrates the results, see also Fig.~3.5 in \cite{Feldman-11}.
We see that the instantaneous Hilbert-based amplitude immediately before the pulse and immediately after it strongly deviates from the actual value. This deviation is of the same order of magnitude as the amplitude's change due to the pulse. We observe similar behavior for the instantaneous phases. 
In summary, we shall interpret the HT-based phase and amplitude with caution in this context. 

We suggest improving the performance of the HT-based inference in the following way. We neglect the instantaneous phases in the intervals $(t_s-\delta_\text{off}, t_s)$ and $(t_s+\delta,t_s+\delta+\delta_\text{off})$, where $t_s$ is the instant of the stimulus's onset and $\delta_\text{off}$ is the offset time. 
Next, we obtain the phase $\vp_{s}$ by extrapolating the linear fit of $\vp(t)$ over the interval 
$(t_s-\delta_\text{off}-\delta_\text{fit},t_s-\delta_\text{off})$ to the instant $t_s$; the parameter $\delta_\text{fit}$ is the length of the fitting interval. Note that fitting requires the unwrapped phase. Similarly, extrapolating the fit over the interval $(t_s+\delta+\delta_\text{off},t_s+\delta+\delta_\text{off}+\delta_\text{fit})$ to $t_s+\delta$ yields $\vp_e$. Moreover, linear fit provides the frequency $\omega$. Hence, we compute $Z_{\cal P}(\vp_s)$.
The amplitude response (ARC) we compute as 
$A(\vp_s)=a_H(t_s+\delta+\delta_\text{off})/a_H(t_s-\delta_\text{off})$.
We emphasize that the algorithm has two parameters, $\delta_\text{off}$ and $\delta_\text{fit}$. 

\begin{figure}
\centerline{\includegraphics[width=0.6\paperwidth]{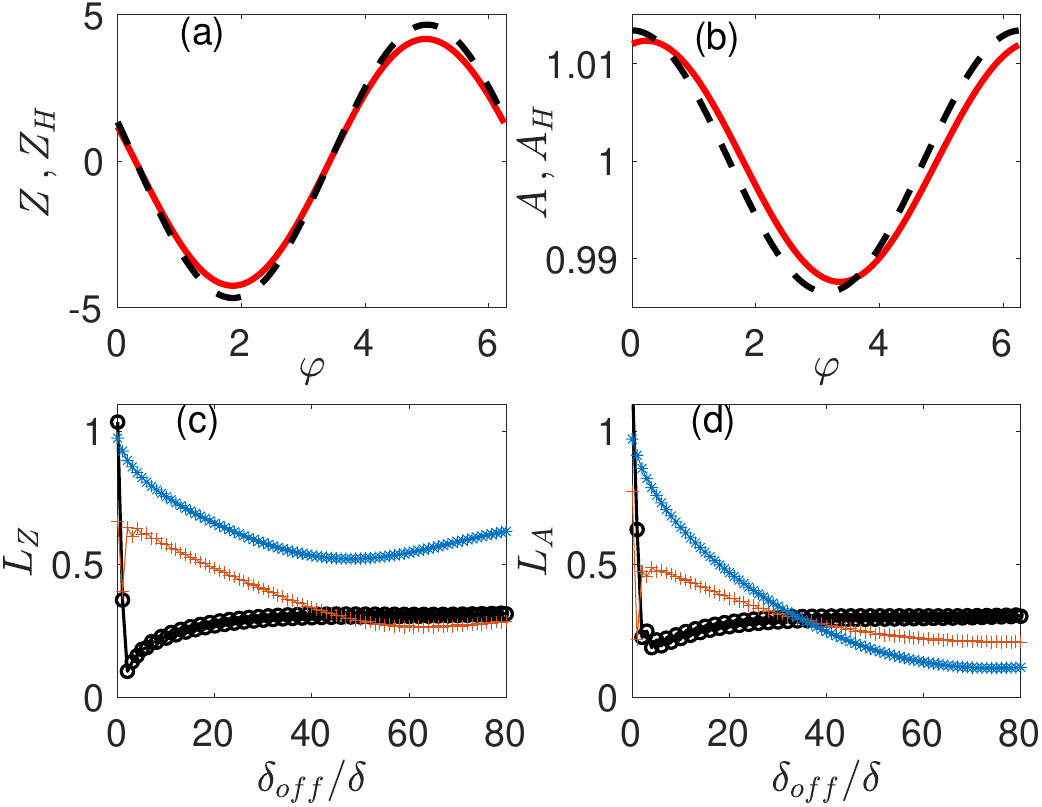}}
\caption{HT-based phase (a) and amplitude (b) responses' inference for the SL oscillator. 
Dashed black and solid red curves show the theoretical and inferred curves, respectively, for $\beta=0$. Parameter $\beta$ describes how the stimulation enters Eqs.~(\ref{eq:SL_cartesian}). Panels (c,d) show the inference errors as a function of the offset parameter $\delta_\text{off}$ for $\beta=0$ (black circles),  $\beta=\pi/4$ (red crosses), and  $\beta=\pi/2$ (blue stars). An optimal choice of $\delta_\text{off}$ ensures successful inference shown in (a,b). However, the optimization requires knowledge of the investigated system, making the HT-based technique of limited use in practical applications. 
}
    \label{fig:HT2}
\end{figure}

We demonstrate the performance of the described algorithm perturbing the SL system (\ref{eq:SL_cartesian}) by rectangular pulses, taking the inter-pulse interval incommensurate with the natural period $T$ -- in this way, we ensure that stimulation occurs at different phases. We set the observable $s(t)=x(t)$ and first choose $\beta=0$.  Other parameters are the same as in the illustration of the standard and sine-fitting techniques. We illustrate the results in Fig.~\ref{fig:HT2}a,b. Here, we show the theoretical curves for $Z(\vp)$, $A(\vp)$ according to Eqs.~(\ref{eq:SL_IRC},\ref{eq:SL_ARC}) along with inferred characteristics $Z_H(\vp)$, $A_H(\vp)$. To compute the latter, we choose $\delta_\text{fit}=50\delta$ and $\delta_\text{off}=2\delta$ and perform an 8th-order Fourier fit of obtained points. 
Numerical tests show that although the results are not very sensitive to the choice of  $\delta_\text{fit}$, the choice of $\delta_\text{off}$ is crucial. We demonstrate this in Fig.~\ref{fig:HT2}c,d by showing the dependence of the error of inference on $\delta_\text{off}$.
We compute this error according to Eq.~\eqref{eq:errorL}.
We see, that a proper choice of the offset $\delta_\text{off}$ essentially affects the inference; the reasonable results in Fig.~\ref{fig:HT2}a,b are due to the optimal value $\delta_\text{off}/\delta=2$. Unfortunately, we do not see any practical way to choose $\delta_\text{off}$ when the true curve is unknown~\footnote{For PRC estimation, a possible approach would be computing the error measure $E_Z$ according to Eq.~(\ref{eq:Z_error}) 
as a function of $\delta_\text{off}$ and searching for the minimum.
However, this approach does not apply to ARC estimation since the Hilbert technique does not yield the Floquet exponent. While we can use the same $\delta_\text{off}$ value to infer both curves, there is no guarantee that the optimal value is the same. }. 
Moreover, the results strongly depend on how the stimulation enters the system's equations. 
To demonstrate this, we plot in Fig.~\ref{fig:HT2}c,d the corresponding curves for $\beta=\pi/4$ and $\beta=\pi/2$. We see that, generally, the error of inference is not small. 
Summarizing this example, we say that the HT-based phase and amplitude response inference generally yields imprecise results. Though one can use this approach to obtain some empirical measures of the response, see~\cite{Duchet_et_al-20}, these inference results may be loosely related to the theory.  

Additionally, we mention that HT serves merely as a signal embedding technique, yielding the angle variable or protophase; see \cite{Kralemann_et_al-07,Kralemann_et_al-08} for a discussion. A protophase generally depends on the embedding; so, e.g., the Hilbert-based protophase does not necessarily coincide with the angle variable in the $x,y$ plane. Although protophases and the true (asymptotic) phase provide the same average frequencies, they generally differ microscopically, i.e., on a time scale smaller than the period. This difference is due to the non-uniform rotational velocity of the protophase. 

\subsubsection{Estimating phase and amplitude using a virtual auxiliary oscillator}\label{sec:auxiliary_osc}

Finally, we adapted the technique \cite{Rosenblum-Pikovsky-Kuehn-Busch-21} for real-time estimation of phase and amplitude for our purpose.
This technique exploits two virtual linear oscillators - one for phase and one for amplitude determination - to yield a causal estimation. Namely, we use the signal $s(t)$ as an input to a damped oscillator
$\ddot x+\alpha_{a,\vp} \dot x +\eta^2 x=s(t).$
We choose the oscillators' frequencies $\eta$ to be much larger than the characteristic frequency $\nu$ of $s(t)$ so that the systems are far from resonance and the response weakly depends on $\nu$. Next, we take the damping parameters $\alpha_\vp$ and $\alpha_a$ for the phase and amplitude measurement, to ensure a simple relation between the phase $\vp(t)=\arctan(-\dot x/\nu x)$ and amplitude $\sqrt{x^2+(\dot x/\nu)^2}$ of the forced oscillation and those of the investigated signal. The implementation is simple and boils down to the numerical integration of the linear oscillator's equation driven by a signal given at discrete time points.  
For details, see Appendix~\ref{app:linosc}.

\begin{figure}
\centerline{\includegraphics[width=0.5\paperwidth]{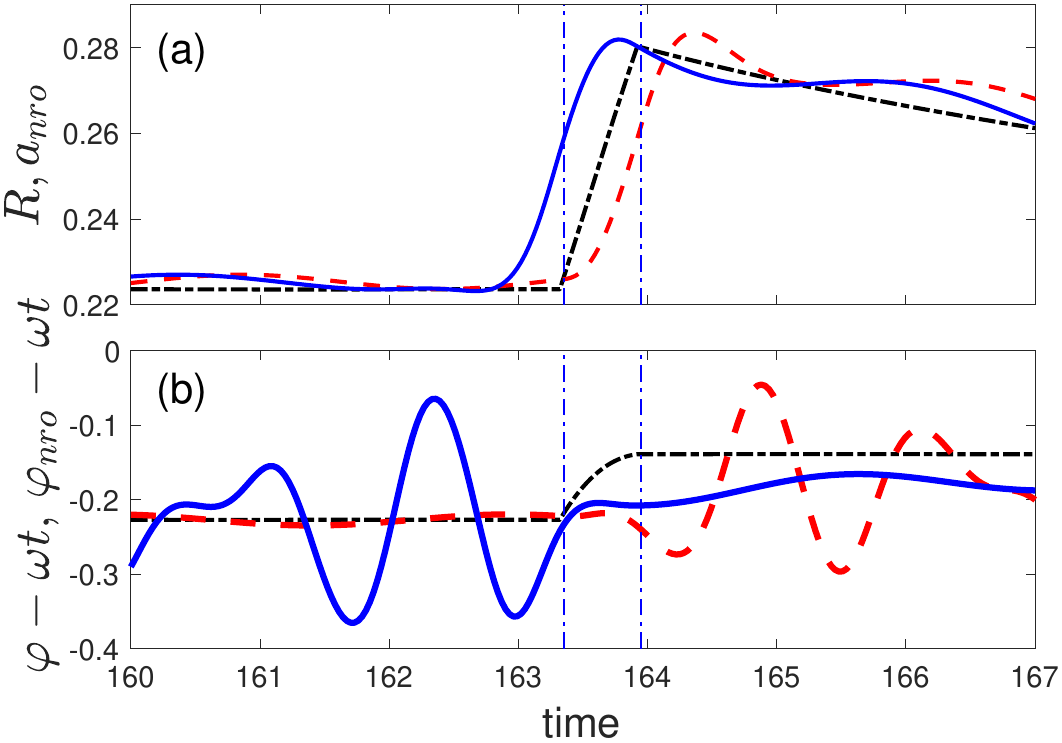}}
\caption{Amplitude and phase of the perturbed SL oscillator, obtained via the linear non-resonant oscillator technique. 
(a) Black dashed-dotted curve shows the time dependence of the amplitude variable
$R=\sqrt{x^2+y^2}$ of the SL system, and the red dashed, and solid blue curves show the instantaneous amplitude $a_\text{nro}$, where the subscript stands for the non-resonant oscillator.
Note that dashed and solid lines correspond to time series $s_k$ and $\hat s_k$, respectively.
(b) Here, the black dashed-dotted curve shows the evolution of the oscillator's phase $\vp$, while the red dashed and solid blue curves illustrate the estimated phase $\vp_\text{nro}$, computed from $s_k$ and $\hat s_k$, respectively (for both phases, we subtract the average growth with the frequency $\w$).
The vertical lines in all panels indicate the beginning and end of the rectangular pulse. We see that the amplitude computed forward in time yields a good estimate of the amplitude immediately before the pulse. Similarly, the amplitude computed backward in time reasonably traces the amplitude after the pulse. For the phases, the estimation is less successful (at least for this particular pulse). 
}
    \label{fig:amp_phase_flip}
\end{figure}

We expect this causal approach to yield a precise estimation of the amplitude and phase {\em before} the stimulus. Immediately after the stimulus, the estimation is poor due to transients. Indeed, the approach implies that the oscillation with the oscillator's frequency $\eta$ decays, and only the oscillation with the frequency of the input $s(t)$ remains. We suggest the following solution to this problem. We compute the amplitude/phase twice, first for the original signal and then for the time-inverted one. Namely, we flip the time series $s_k$ so that the time ``runs'' backward, i.e., from the $N$-point time series $s_k$,  $k=1,2,\ldots,N$, we construct the flipped series $\hat s_1=s_N$, $\hat s_2=s_{N-1}$, \ldots, $\hat s_N=s_1$. Next, we compute the amplitude of $\hat s_k$ via the same algorithm and flip it in time to obtain the flipped amplitude denoted by $\hat a(t)$.  
$\hat a(t)$ is a precise estimation for the time interval after the stimulus, while the transients corrupt the estimation before the stimulus. However, since the phase changes its sign with the time inversion, we must reflect the obtained phase, $\hat \vp(t)\to- \hat \vp(t)$. 
We illustrate this algorithm in Fig.~\ref{fig:amp_phase_flip}; parameters of the virtual oscillators are $\eta=5$, $\alpha_a=6$, $\alpha_\vp=0.2$.

\begin{figure}
\centerline{\includegraphics[width=0.5\paperwidth]{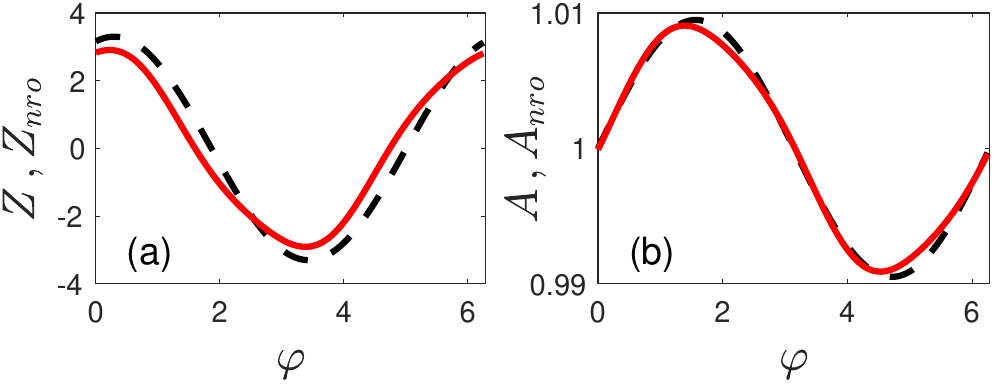}}
\caption{Phase (a) and amplitude (b) responses for 
the SL oscillator, inferred using auxiliary non-resonant linear oscillators.
Dashed black and solid red curves show the theoretical and inferred curves, respectively. We show the results for
$\beta=\pi/2$, where the HT-based technique fails. We recall that parameter $\beta$ describes how the stimulation enters Eqs.~(\ref{eq:SL_cartesian}).
}
    \label{fig:nro_curves}
\end{figure}

A remark is in order. The linear oscillator used to compute the phase has a smaller damping parameter than the oscillator for the amplitude estimation. Correspondingly the transients in the phase measurement are essentially longer than in the case of the amplitude measurement, which makes the phase measurement less precise. 
We present the inference's results for the SL model in Fig.~\ref{fig:nro_curves}. Additionally, just like any approach relying on estimating small differences in the angle variable, it generally suffers significant errors because it does not include information on local isochrons.

\subsection{Inferring response curves by reconstructing the first order phase-isostable dynamics: the IPID-1 technique}
\label{sec:phaseiso}

Here, we describe a method for inferring the first-order phase-isostable dynamics from observations by fitting the model Eqs.~(\ref{eq:Winfree},\ref{eq:iso}). We name the method IPID-1 standing for ``Inferring Phase--Isostable Dynamics of order 1''. We carry out the procedure in two steps. First, we infer the PRC and instantaneous phase by adapting the algorithm introduced in Ref~\cite{cestnik_rosenblum_2018} for the case of pulse stimulation. Next, we use the inferred phase to reconstruct the isostable dynamics. 
As in the previous sections, we assume that a scalar signal $s(t)$ and the perturbation $p(t)$ are known. 

\subsubsection{Inference of the phase response}\label{sec:PRC_inf}
The key step in the inference is determining time instants $\tau_i$ corresponding to the same asymptotic phase $\vp$; these instants must be extracted from the observed signal $s(t)$. Sometimes the choice of such events is obvious, e.g., in the case of a spiky signal where spikes indicate the same phase. In general one considers threshold-crossing events, $s(\tau_i) = s_\text{thr}$, choosing one crossing per period, e.g., with an additional condition $\frac{d}{dt} s(\tau_i) > 0$.  The choice of the threshold $s_\text{thr}$ is important. The proper thresholding should closely match the crossing of a local isochron, see Appendix~\ref{app:threshold_crossing} for details. Since isochrons make a full rotation over a limit cycle, at least two such thresholds exist for any scalar signal - we find appropriate thresholds with a direct search, seeing which threshold yields the best fit in terms of the error~\eqref{eq:Z_error}. 

The core of the inference is fitting the Winfree phase equation~\eqref{eq:Winfree} integrated over individual periods, determined as the time interval between two threshold-crossing events, $[\tau_i,\tau_{i+1}]$:
\begin{equation}
    2\pi = \omega (\tau_{i+1}-\tau_i) + \int\limits_{\tau_i}^{\tau_{i+1}} Z(\vp(t)) p (t) \dd t \;.
    \label{eq:Winfree_int}
\end{equation}
The left-hand side equals $2\pi$ due to the definition of events $\tau_i$ having the same phase. On the right-hand side, we approximate the PRC by a finite Fourier series of order $N_F$: $Z(\vp) = \sum_{n=0}^{N_F}\left [  z_n^{\cos} \cos(n\vp) + z_{n}^{\sin} \sin(n\vp)\right ]$. By interchanging the order of integration and summation, we obtain from Eqs.~\eqref{eq:Winfree_int} a linear system for the unknown Fourier coefficients $z_n$ and frequency $\omega$. 
Since we have as many equations~\eqref{eq:Winfree_int} as there are periods of the observed signal, for a sufficiently long data set we have more equations than unknowns.
Thus, we can solve the linear system, e.g., by least-squares minimization. The coefficients $\int \cos(n\vp)p(t) \dd t$,
$\int \sin(n\vp)p(t) \dd t$,
however, cannot be evaluated yet because the phase $\vp(t)$ is not known to us {\it a priori}. We overcome this with an iterative procedure; by first approximating the phase to obtain an approximate solution, and then exploiting this solution to improve the phase estimate~\footnote{Each phase approximation $\vp^{(m)}$ is used to compute the integrals that represent coefficients of the linear system: $\int\cos(n\vp^{(m)})p(t)\dd t$, $\int\sin(n\vp^{(m)})p(t)\dd t$. Then when solving the linear system and obtaining a better PRC approximation $Z^{(m+1)}(\vp)$, it is used to recompute the next approximation of the phase $\vp^{(m+1)}$ by integrating the Winfree Eq.~\eqref{eq:Winfree}}. 

Approximating the phase initially as linearly growing between the events, $\vp^{(0)}(t) = 2\pi \frac{t-\tau_i}{\tau_{i+1}-\tau_i}$, 
we obtain the first-approximation solution of  Eq.~(\ref{eq:Winfree_int}), namely $\omega^{(1)}$ and $Z^{(1)}(\vp)$. 
(The superscripts denote the iteration number.) 
Then, we obtain the next approximation of the phase by integrating Winfree Eq.~\eqref{eq:Winfree} between events. In general, the $m^\text{th}$-order approximation for the phase is obtained as:
\begin{equation}
    \vp^{(m)}(t) = \int_{\tau_i}^t \left [  \omega^{(m)} + Z^{(m)}(\vp^{(m)}(t')) \right ] \dd t'\;.
    \label{eq:phase_approx}
\end{equation}
Estimated in this way, the phase at the end of a period generally differs from $2\pi$: $\Phi_i^{(m)} = \lim\limits_{t\uparrow \tau_{i+1}} \vp^{(m)}(t) \neq 2\pi$. Thus, we additionally re-scale the phase with the factor $\Phi_i^{(m)}$ to ensure $\vp^{(m)}(\tau_{i+1})\equiv 2\pi$. 
As a result, the approximations gradually improve through iterations.
Estimated phases at the end of periods $\Phi_i$ indicate how well our inference fits the observations, see error measure~\eqref{eq:Z_error}. For each iteration $(m)$ we can compute the error and monitor the convergence. 
For further details of the technique, we refer to Ref.~\cite{cestnik_rosenblum_2018}. 

\subsubsection{Inference of the isostable variable response}\label{sec:IRC_inf}
This section extends our approach to cover the isostable dynamics reconstruction. We still need the signal $s(t)$ and perturbation $p(t)$, and since we have already inferred the PRC, we also have the instantaneous phase $\vp(t)$. 

Like the PRC technique, the isostable inference relies on time events of equal phase $\tau_i$, and  additionally, on estimating the isostable variable at those events, $\psi_i \equiv \psi(\tau_i)$. The time events $\tau_i$ are straightforwardly obtained from the instantaneous phase $\vp(t)$~\footnote{Note that the events $\tau_i$ used for isostable inference generally differ from those used for the PRC reconstruction, though we keep the same notation. Practically, to determine 
$\tau_i$ we use linear interpolation of the monotonically growing function $\vp(t)$ given in a discrete set of points.}, 
while the isostable variable needs to be estimated from the observed signal $s(t)$. The IRC function describes a linear response and is operable in the close vicinity of the limit cycle, where straight lines can approximate isochrons. Thus in the first-order approximation, the isostable variable linearly depends on signal $s(t)$:
\begin{equation}
    \psi_i = c\ (s(\tau_i)-s_0) + \mathcal{O}((s(\tau_i)-s_0)^2)\;,
    \label{eq:psi_approx}
\end{equation}
where factors $c$ and $s_0$ are generally phase-dependent. However, since all crossing events $\tau_i$ correspond to the same phase, $c$ and $s_0$ are constant in this context. Additionally, the isostable variable is inherently determined up to a constant factor. Therefore,  without loss of generality, we can set $c \equiv 1$. 
Note that approximation via Eq.~\eqref{eq:psi_approx} is also justified for high-dimensional systems given if one Floquet exponent is significantly smaller in absolute value than the rest, i.e., there is a slow manifold (see discussion in Appendix~\ref{app:phipsi}). 

We now consider the isostable dynamics~\eqref{eq:iso} integrated over periods determined by the events of the same phase $[\tau_i,\tau_{i+1}]$:
\begin{equation}
    \psi_{i+1}-\psi_i = \kappa \int_{\tau_i}^{\tau_{i+1}}\psi(t) \dd t + \int_{\tau_i}^{\tau_{i+1}} I(\vp)p(t) \dd t \;.
    \label{eq:iso_int}
\end{equation}
This system of equations is similar to Eq.~\eqref{eq:Winfree_int} in that it can be approximated with a linear system by expanding the unknown function in a Fourier series, 
$I(\vp) = \sum_{n=0}^{N_F}\left [  u_{n}^{\cos} \cos(n\vp) + u_{n}^{\sin} \sin(n\vp)\right ]$, and interchanging the order of integration and summation. 
Furthermore, since we know the phase as a function of time, the $i$-dependent factors of Fourier coefficients can be computed directly. 
Using approximation~\eqref{eq:psi_approx} we evaluate the left-hand side as $\psi_{i+1}-\psi_i = s(\tau_{i+1})-s(\tau_i)$
because the still unknown constant $s_0$ cancels in the difference. The integral $\int \psi(t) \dd t$ is challenging since it includes a time-dependent isostable variable, which we do not have.

There are two issues with evaluating the isostable variable integral $\int_{\tau_i}^{\tau_{i+1}}\psi(t) \dd t$: (i) we do not know the constant $s_0$ {\it a priori}, and (ii) we want to use approximation~\eqref{eq:psi_approx} only in discrete events $\tau_i$ (otherwise, we would have to consider $c$ and $s_0$ as phase-dependent). We tackle the first issue by splitting the integral in two: \begin{equation}
    \kappa \int_{\tau_i}^{\tau_{i+1}} \psi(t) \dd t = - \kappa s_0 (\tau_{i+1}-\tau_i) + \kappa \int_{\tau_i}^{\tau_{i+1}} (\psi(t)+s_0) \dd t\;.
\end{equation}
In this representation $\kappa s_0$ becomes just another unknown variable that will be determined while solving the linear system $\eqref{eq:iso_int}$, which together with determining $\kappa$ allows expressing $s_0$ as the ratio $s_0 = \frac{\kappa s_0}{\kappa}$. The remaining integral involves the continuous quantity $\psi(t)+s_0$, which coincides with our observable $s(t)$ at events $\tau_i$ (generally this quantity differs from the signal for $t \neq \tau_i$: $s(t) \neq \psi(t)+s_0$). 
The property $s(\tau_i) = \psi(\tau_i)+s_0$ also helps us resolve the second issue of only using discrete values $s(\tau_i)$ to estimate the integral of the continuous isostable variable. Just as in the phase response method, we approach this iteratively. First, we linearly interpolate $\psi(t)+s_0$ between known events so that we can evaluate the last term: $\int_{\tau_i}^{\tau_{i+1}} (\psi(t)+s_0)\dd t \approx [s(\tau_i)+s(\tau_{i+1})](\tau_{i+1}-\tau_i)/2$. Then we solve the linear system Eq.~(\ref{eq:iso_int}) and obtain the first-approximation solution $s_0^{(1)}$, $\kappa^{(1)}$, $I^{(1)}(\vp)$. Next, we use this solution to improve the estimate of the continuous isostable variable by integrating the underlying dynamical equation~\eqref{eq:iso}. In general, from the  approximate solution of order $m$, we obtain the subsequent isostable variable estimation via integration~\footnote{Since the forcing term is just a function of time, the isostable dynamics can be integrated via variation of constant, in which case the time integral can be computed explicitly: $\psi^{(m)}(t) = (s(\tau_i)-s_0^{(m)})e^{\kappa^{(m)}(t-\tau_i)} + \int_{\tau_i}^t I^{(m)}(\vp(t'))p(t')e^{\kappa^{(m)}(t-t')} \dd t'$}:
\begin{equation}
    \psi^{(m)}(t) = s(\tau_i)-s_0^{(m)} + \int_{\tau_i}^t [\kappa^{(m)} \psi^{(m)}(t') + I^{(m)}(\vp)p(t') ] \dd t' \;.
    \label{eq:integrate_iso}
\end{equation}
Estimated in this way, the isostable variable $\psi^{(m)}$ starts the interval $[\tau_i,\tau_{i+1}]$ as a direct estimation from the signal: $s(\tau_i)-s_0^{(m)}$, but further along the interval it can deviate due to the approximation in integrands $\kappa^{(m)},I^{(m)}$. Thus, at the end of the interval this quantity does not exactly correspond to $s(\tau_{i+1})-s_0^{(m)}$, which is the value that we take to start the next interval $[\tau_{i+1},\tau_{i+2}]$. The estimated time series of the isostable variable obtained in this way is thus discontinuous at the events $\tau_i$. 
If the inference were perfect, this discontinuity would disappear, which means we can use the magnitude of the discontinuity to quantify the quality of the model as follows. 
Let us denote the isostable variable estimated at the end of a period as $\Psi_i^{(m)} = \lim\limits_{t\uparrow \tau_{i+1}} \psi^{(m)}(t) \neq s(\tau_{i+1})-s_0^{(m)}$. We define the error of the fit as the standard deviation of the difference: 
\begin{equation}
    E_I^{(m)} = \langle (\Psi_i^{(m)}-(s(\tau_{i+1})-s_0^{(m)}))^2 \rangle^{1/2}\;,
    \label{eq:I_error}
\end{equation}
Just like the PRC error was compared to the irregularity of inter-event intervals, see Eq.~\eqref{eq:Z0_error}, this error should be compared to the irregularity of the isostable variable at the events:
\begin{equation}
    E_{I0} = \langle (s(\tau_{i})-\langle s(\tau_i) \rangle )^2 \rangle^{1/2}\;.
    \label{eq:I0_error}
\end{equation}
The entire procedure of inferring the isostable variable response is condensed into step-by-step instructions in Appendix~\ref{sec:step_by_step}. 

\subsubsection{Performance of the IPID-1 phase-isostable reconstruction}\label{sec:phase-iso-rec}
\begin{figure}
    \centering
    \includegraphics[width=0.8\paperwidth]{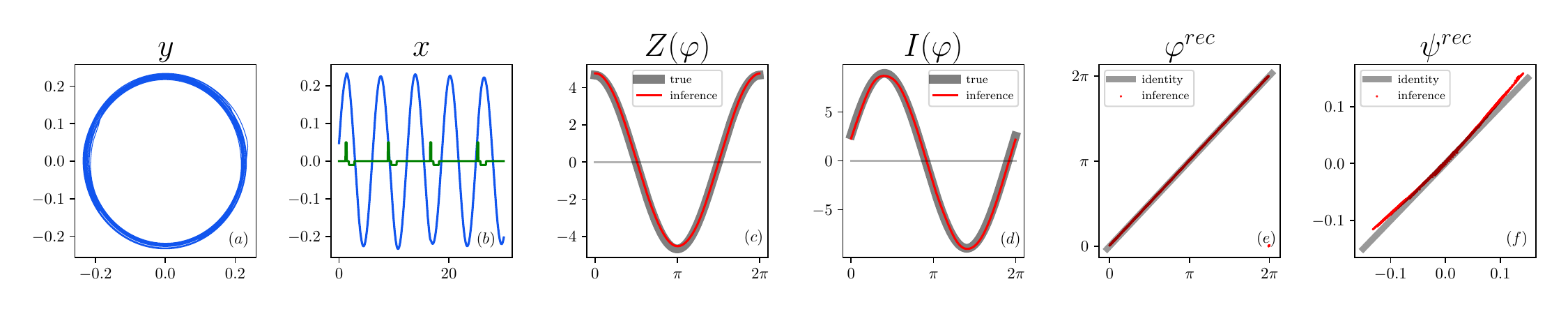}\\
    \includegraphics[width=0.8\paperwidth]{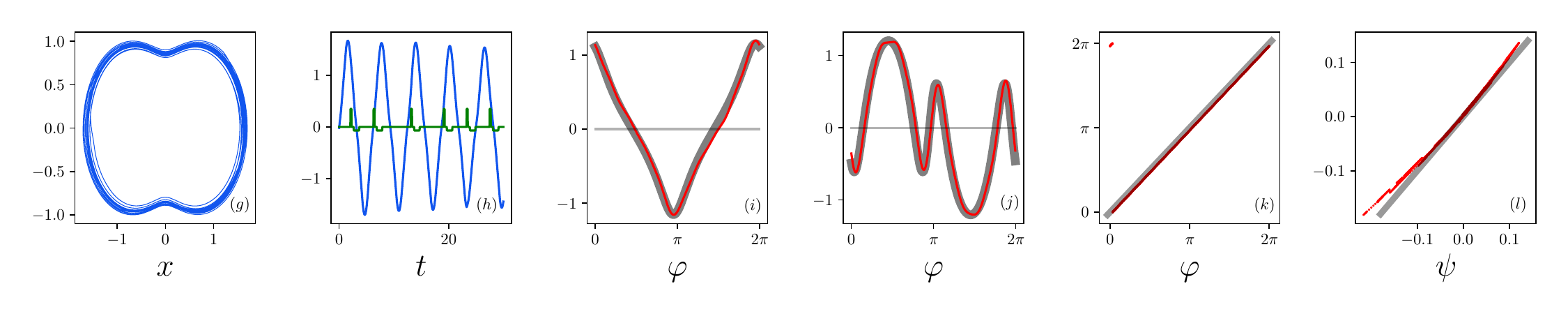}\\
    \caption{Results of PRC and IRC inference  with the IPID-1 method from Sec.~\ref{sec:phaseiso} for the two test oscillators: Stuart-Landau \eqref{eq:SL_cartesian} (top row) and its generalization, Eq.~\eqref{eq:infosc} (bottom row). From left to right: (a,g) trajectory in state space, (b,h) an epoch of the observed signal and pulsatile forcing (blue and green), (c,i) phase response curve, (d,j) isostable response curve, (e,k) asymptotic phase comparison, (f,l) isostable amplitude comparison. We put the labels for the vertical axes on the top of the plots to save space.  }
    \label{fig:inference_example}
\end{figure}

We test the performance of the introduced IPID-1 method on the two example oscillators~\eqref{eq:SL_cartesian} and \eqref{eq:infosc}. We infer the PRC $Z(\vp)$ and IRC $I(\vp)$ with the approaches explained in Sections~\ref{sec:PRC_inf} and \ref{sec:IRC_inf} respectively. The results can be seen in Fig.~\ref{fig:inference_example}. Parameters are the same as used in Sec.~\ref{sec:standard_technique} and Fig.~\ref{fig:prc_standard}. We stimulate with a bipolar charge-balanced pulse with a period of positive stimulation lasting 0.2, a period of no stimulation lasting 0.4, and a period of negative stimulation lasting 1.0 (pulse shape can be seen compared to the signal in Fig.~\ref{fig:inference_example}b,h). We used a relatively long time series (1500 periods), and there was no noise or other unknown inputs (see Appendix~\ref{app:forcing_gen} for details). Additionally, we determined events of equal phase with the optimal threshold~\footnote{The optimal threshold corresponds to thresholding with a Poincar\'e section that is tangential to a local isochron, see Fig.~\ref{fig:thresholding}. We find it by performing a direct search over a reasonable range of threshold values and choosing the one that corresponds to the lowest error value~\eqref{eq:Z_error}, as explained in the method paper~\cite{cestnik_rosenblum_2018}.} 
by performing a direct search over possible threshold values, minimizing the error~\eqref{eq:Z_error}. As a result, the inferred curves (red) accurately reflect the true ones (thick gray). The slight deviations are mostly due to using a finite Fourier representation ($N_F = 10$). As a byproduct of the inference, we also obtain the asymptotic phase and isostable amplitude as functions of time, and plot them against their true counterparts in Fig.~\ref{fig:inference_example}e,f,k,l. The obtained isostable time series is also a good representation of the signal envelope if shifted to match the signal maxima, as shown in a later Figure.~\ref{fig:arc_comparison}. 

\subsection{Inference in the presence of noise}
\begin{figure}
\centerline{\includegraphics[width=0.5\paperwidth]{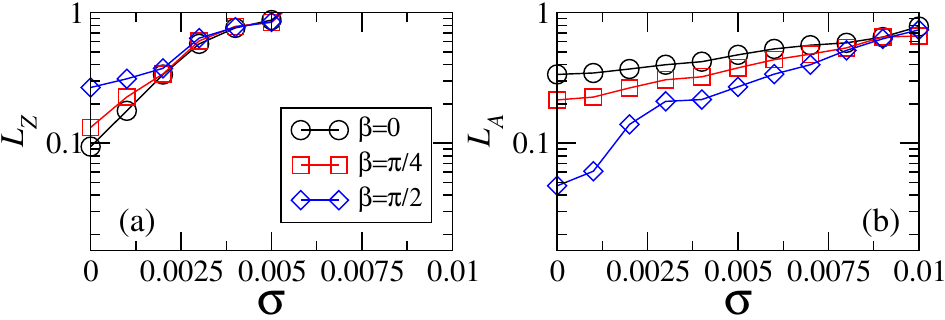}}
\caption{Error~\eqref{eq:errorL} of PRC (a) and ARC (b) inference using an auxiliary oscillator as explained in Sec.~\ref{sec:auxiliary_osc}. The error is measured for different noise strengths $\sigma$. A single realization of noise was used. 
}
    \label{fig:nro_beta}
\end{figure}

Real-world oscillators are inevitably noisy, and therefore, we test the performance of introduced techniques in the presence of dynamical noise. 
For this purpose, we simulate the same two systems~\eqref{eq:SL_cartesian} and \eqref{eq:infosc} with different strengths $\sigma$ of Gaussian white noise.
We start with the SL oscillator and test the sine-fitting, Hilbert-based, and auxiliary oscillator techniques; again we use the unipolar pulses. In Fig.~\ref{fig:nro_beta} we illustrate the performance of the latter method. (Since the sine-fitting and Hilbert-based techniques are less efficient for noise-free systems, we do not illustrate them here.) We show the inference errors as a function of the noise intensity $\sigma$, for three different values of $\beta$, cf. Eq.~(\ref{eq:SL_cartesian}).
As expected, for this method the PRC inference is more sensitive to noise than the ARC reconstruction. 

\begin{figure}
    \centering
    \includegraphics[width=0.8\paperwidth]{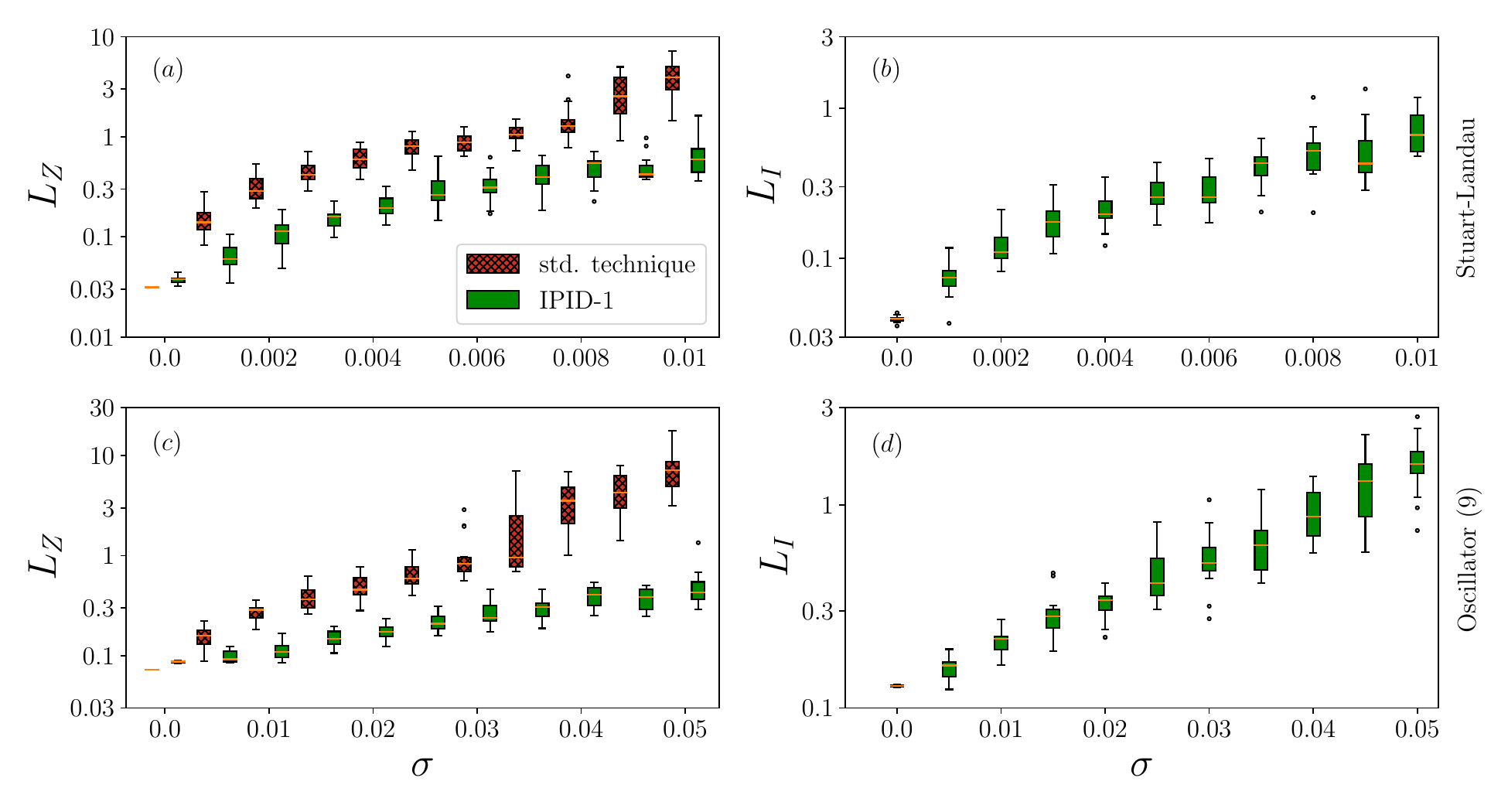}
    \caption{Performance of the standard technique~\ref{sec:standard_technique} (red, crossed) and introduced IPID-1 method~\ref{sec:phaseiso} (green, plain) in relation to dynamical Gaussian white noise strength $\sigma$ (the red crossed and green plain box plots are slightly shifted horizontally to the left and to the right, respectively, to avoid overlap). Two test oscillators were used, Stuart-Landau~\eqref{eq:SL_cartesian} (a,b) and the modification~\eqref{eq:infosc} (c,d). 
    Panels (a,c) and (b,d) show the error measures according to Eq.~(\ref{eq:errorL}) for the phase response $Z(\vp)$ and 
    the amplitude response $I(\vp)$, respectively.
    For each value of noise strength $\sigma$, several trajectory realizations were simulated, each realization yielding one error measurement. The $y$-axis is logarithmic, and the measurements for the same $\sigma$ are represented with standard box plots. Parameters of the oscillators and the perturbing pulses are the same as used before in Sec.~\ref{sec:standard_technique} and Figs.~\ref{fig:prc_standard} and~\ref{fig:inference_example}.   
    }
    \label{fig:inference_noise}
\end{figure}

Next, we proceed with testing the most promising IPID-1 technique, based on the phase-isostable dynamics reconstruction. 
We keep the forcing pulse action constant, and then for each $\sigma$ generate several trajectory realizations, and for each one, we compute the $L$ error measure using Eq.~\eqref{eq:errorL}.
These errors are then depicted with standard box plots in Fig.~\ref{fig:inference_noise}a,b. For comparison, we also show the results for the standard technique (PRC only). We conclude that the PRC inference with the technique introduced in Section~\ref{sec:PRC_inf} is most stable to noise; in particular, it outperforms the standard approach. The direct comparison of the amplitude response reconstruction is not that easy since the auxiliary oscillator approach yields the empirical amplitude and, correspondingly, the ARC, while the isostable reconstruction provides the IRC. However, the results presented in Fig.~\ref{fig:nro_beta}b and 
Fig.~\ref{fig:inference_noise}b indicate approximately equal performance.    
We underline that stronger forcing would imply lower errors in the strong noise regime since the forcing-to-noise ratio would increase. On the other hand, the errors in the weak noise regime would increase since the linear approximation works worse for strongly forced systems. We also mention that the standard technique (red) has stronger limitations on the observations, namely, the pulses have to be rare (one pulse per several periods), which is why in the case of no noise, the standard technique has a marginally smaller error. For our technique, we considered a more realistic case of more than one pulse per period on average. It means our technique requires a much shorter time series, see Appendix section~\ref{app:forcing_gen} for details.  

Finally, we test our phase-isostable reconstruction technique on the complicated case of the modified SL oscillator~\eqref{eq:infosc} that provides a non-sinusoidal observable, cf. Fig~\ref{fig:infinity_oscillator}b. The results in Fig.~\ref{fig:inference_noise}c,d confirm the advantage of this technique, namely, it is not restricted to sinusoidal signals. Moreover, the inference errors $L_Z$, $L_I$ for the modified SL are not much different from those for the standard SL that provides a sinusoidal signal. The results in panel (c) clearly demonstrate that our technique is more stable with respect to noise than the standard one.

\subsection{Response of a high-dimensional system}
As a final test we apply the studied and newly-developed techniques to a signal generated 
by a globally coupled oscillatory ensemble, see Eqs.~\eqref{eq:bvdp}.
The individual units are governed by the Bonhoeffer–van der Pol equations.
This system can be treated as a simple model of neuronal rhythmical activity.

\begin{figure}
    \centering
    \includegraphics[width=0.75\paperwidth]{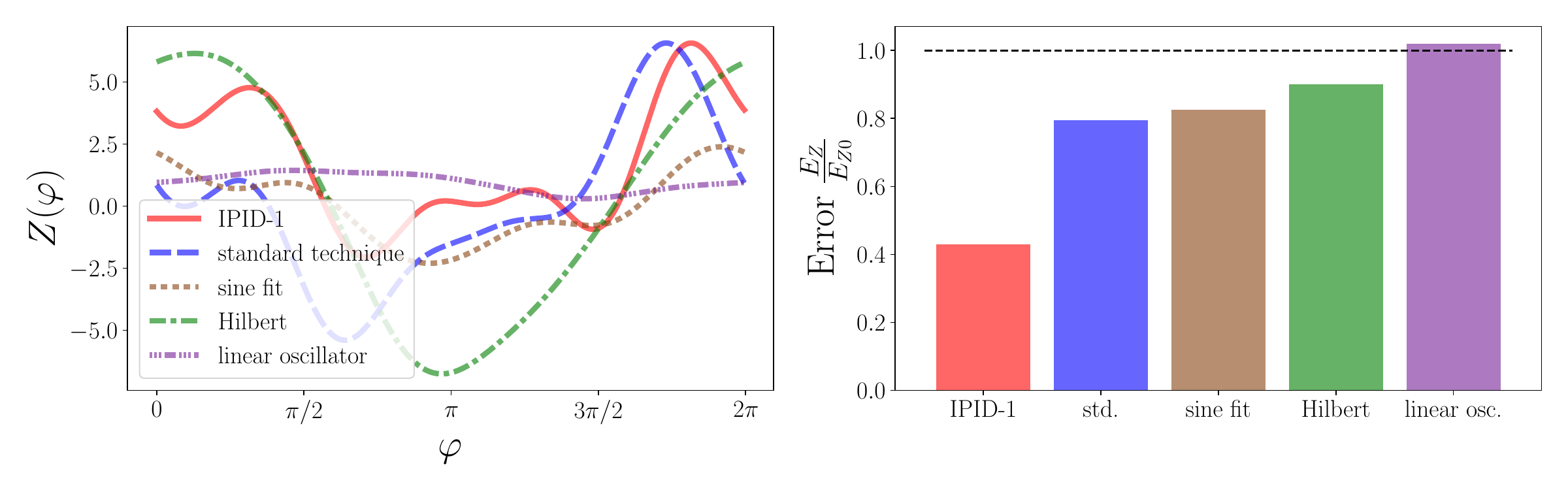}
    \caption{Application of all phase response methods on a high dimensional example of a chaotic oscillatory ensemble~\eqref{eq:bvdp}. (a) Inferred PRCs and  (b) the ratio of errors \eqref{eq:Z_error} and \eqref{eq:Z0_error}, 
    representing the goodness of the inferred model (ratio 0 would correspond to a perfect model, and ratio one is as good as considering no response). }
    \label{fig:ensemble_prcs}
\end{figure}

We consider a large number of oscillators and choose the parameters such that the system exhibits weak collective chaos, see test oscillators Sec.~\ref{sec:test} for more details on the system and Fig.~\ref{fig:ensemb_traj} for a depiction of the mean-field orbit. 
We assume that we observe collective dynamics. 
This example is a hard test. 
In this case, not only is the ground truth phase model unknown to us, but since the system is chaotic, we know that there exists no phase description that can exactly describe the local deviations~\cite{chaotic_schwabedal}. We, therefore, estimate the goodness of inference solely based on how well the inferred model reproduces observations; namely, we consider the ratio of errors~\eqref{eq:Z_error} and \eqref{eq:Z0_error}: $E_Z/E_{Z0}$. Our IPID-1 method performs best and yields an error ratio of less than half, while other methods perform rather poorly. Figure~\ref{fig:ensemble_prcs} shows all the corresponding phase response curves and the values of errors in a bar plot. We inferred the empirical PRCs using the mean field in the $x$ variable as our observable, $s=X$, and exploiting bipolar pulses as before. Next, we deconvolved all curves to obtain the theoretical phase response curves. We also performed the inference on a different observable $s=X+2Y$ of the same system and obtained similar results.  

Only three techniques can infer the amplitude response: the Hilbert-based,  auxiliary oscillator-based, and IPID-1. The first two methods yield the effective ARC. In contrast, the IPID-1technique estimates the infinitesimal isostable curve $I(\vp)$. Hence, for comparison, we recompute $I(\vp)$ into the ARC in the following way: first, using the infinitesimal curve $I(\vp)$ and pulse shape $\mathcal{P}(t)$ we evaluate the isostable shift $\Delta \psi$. Then we relate it to the ARC. For a particular phase $\vp^*$,  we integrate the first-order dynamics~(\ref{eq:Winfree},\ref{eq:iso}) with the considered pulse shape $\mathcal{P}(t)$ to obtain phase as a function of time, $\vp(t)$, by solving the Winfree equation with the initial condition $\vp(t=0) = \vp^*$, and then compute the effective isostable shift as:
\begin{equation}
\Delta \psi(\vp^*) = \int\limits_{0}^\delta (\kappa \psi(t) + I(\vp(t))\mathcal{P}(t))\dd t\;.
\end{equation}
Next, we have to relate the isostable variable $\psi$ to amplitude $a$. As already mentioned, locally, the isostable linearly depends on the distance from the limit cycle, see Eq.~(\ref{eq:psi_approx}): $\psi \approx c(\vp) (R-R_0(\vp))$.  Here $R$ represents the distance from the origin, while $c(\vp)$ and $R_0(\vp)$ are phase-dependent and hold the information of the isostable structure and parametrization of the limit cycle, respectively. If amplitude is simply defined as $a=R$, then $a= R_0(\vp) + \psi/c(\vp)$. Recalling the definition of ARC as a ratio of amplitudes before and after a pulse, we write:\begin{equation}
A(\vp)= \frac{a_e}{a_s} = 1 + \frac{\Delta\psi(\vp)}{c(\vp)R_0(\vp)}\;.
\end{equation}
We cannot estimate $c(\vp)$ nor $R_0(\vp)$ from a scalar signal -  therefore, here we approximate them by constants.  We choose $c=1$ since this value was used in the inference, and for $R_0$ we use the maximal value of the signal: $R_0 = s_\text{max}$~\footnote{Ignoring the phase dependence of $c(\vp)$ is similar to ignoring the information of local isochrons in the phase response, as most techniques do, e.g., the Hilbert-based one.}. We stress that difficulties of recomputing IRC into ARC are due to an {\it ad hoc} definition of the amplitude. Indeed, defined as the distance from the origin, it depends on the projection space (depends on both the observable and embedding technique). In contrast, $I(\vp)$ is an invariant characteristic of a limit cycle. 

\begin{figure}
    \centering
    \includegraphics[width=0.7\paperwidth]{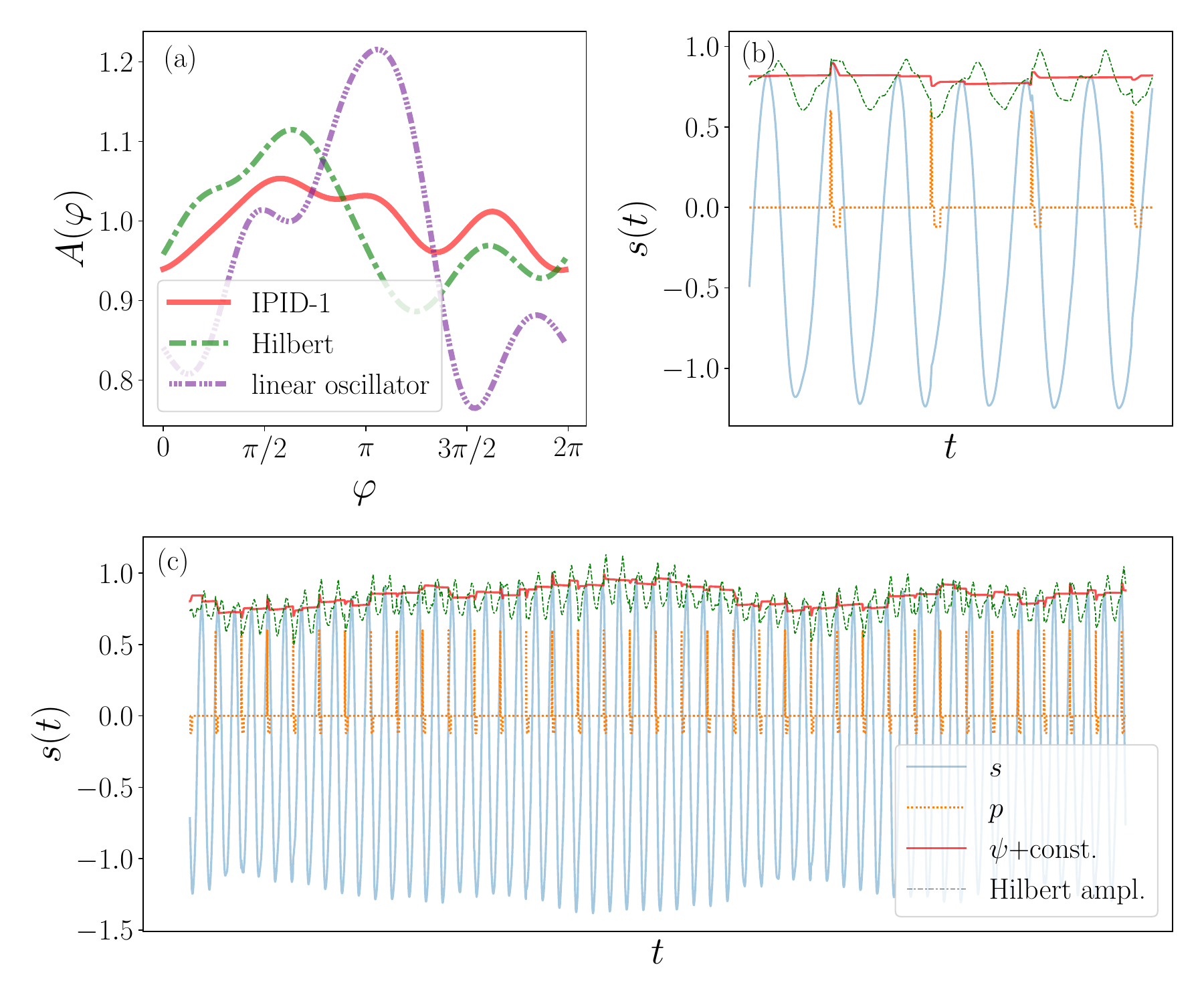}
    \caption{Inferred amplitude response (a) and the signal envelope (b,c) for the high-dimensional chaotic ensemble~\eqref{eq:bvdp}. Three methods were used for inferring the amplitude response: Hilbert-based, virtual oscillator-based, and IPID-1. Since IPID-1 yields the theoretical IRC, we recompute it into ARC, see text. 
    In (b,c) we show two signal envelopes: the Hilbert amplitude (green dash-dotted) and shifted isostable amplitude $\psi$ (full red line). Plot (b) shows the behavior on a short time scale (few periods), and plot (c) shows it on a long timescale (dozens of periods).  }
    \label{fig:arc_comparison}
\end{figure}

We compare the curves in Fig.~\ref{fig:arc_comparison}a. Similarly to the characterization of the phase response, we estimate the goodness of the IRC inference by quantifying how well the dynamical model fits the observation. Namely, we compute the ratio of errors \eqref{eq:I_error} and \eqref{eq:I0_error}. We obtain $E_I/E_{I0}=0.28$, which indicates a decent inference (we remind that this measure is 0 for a perfect fit and of order one for a completely wrong curve). Note that the two other techniques do not yield dynamical equations; therefore, no errors were computed.
As a byproduct, our technique IPID-1 also yields an estimation of $\psi$ as a function of time. In Fig.~\ref{fig:arc_comparison}b,c, we demonstrate 
that the isostable variable provides a good envelope if shifted to match the signal's maxima (or minima). We depict such an envelope along with the Hilbert amplitude as the commonly used alternative. Notice how Hilbert amplitude significantly fluctuates on the timescale of one period, while the isostable envelope changes mainly in times of stimulation and otherwise slowly follows the signal. 

\section{Discussion and conclusions}
\label{sec:discus}

This paper addressed the problem of inferring an oscillator's phase and amplitude response from observations using test stimulation. Below, we discuss these two tasks of phase and amplitude inference separately. 

In addition to critically testing several techniques presented in the literature, we also developed a new method for fitting first-order phase-isostable dynamics to observations, denoted as the IPID-1 method. 
We concentrated on the case where we apply a specially designed stimulation. However, this technique can be exploited for almost any input, as long as we can observe it and it does not entrain the oscillator~\footnote{Another example of a problematic perturbation is periodic stimulation with a frequency faster than the observed oscillator. In the latter case, there is more than one stimulus per period, and the phase of subsequent stimulations $\vp^{*(k)}$ fully depends on the phase of the first one, $\vp^{*(1)}$. This makes distinguishing the phase-shift contributions of individual pulses $Z(\vp^{*(k)})$ impossible.}.

\subsection{Phase response}
PRC determination for noise-free neuron-like oscillators has been known for decades. We here concentrated on noisy signals without well-defined marker events, such as spikes. 
As our first result, we mention the technique for recomputing the empirical PRC obtained in response to stimuli of arbitrary, but known shape, into the theoretical PRC, $Z_{\cal P}(\vp)\to Z(\vp)$. 
This technique can be used alongside with any approach for PRC inference, in particular, it can enhance the applicability of the standard approach. This addition is especially beneficial for analyzing biological systems where some restrictions on the stimulus's shape may apply. 

Next, we suggested a simple approach to inference error estimation. Without knowing the ground truth and using only the observations, we quantify the inferred PRC's accuracy in reproducing the observations. 
Namely, we quantify how much of the signal's variability can be explained by the inferred PRC. Again, this approach can complement any PRC estimation technique. We strongly recommend using this error estimation tool in experiments. Indeed, blind application of an inference technique always provides some response curve, but in the case of noisy or chaotic systems, the obtained curve may not describe the underlying dynamics.   

We critically considered previously developed techniques that were  introduced without being tested on examples where the ground truth
is known. We performed tests and highlighted the potential weaknesses and drawbacks. In particular, we suggested improving the inference approach based on the popular analytical signal technique exploiting the HT by ignoring the data points immediately before and after stimuli. Furthermore, we compared the performance of the known and newly developed techniques in the noise-free and noisy cases using specially designed test data with the known ground truth. We tested the dependence of the results on the noise level and the observable used for data analysis. In particular, we have shown that the HT-based technique is generally unreliable. As a result, we have demonstrated the essential advantage of our IPID-1 approach based on the direct reconstruction of the Winfree equation: it is not restricted to sine-like signals, it immediately provides the theoretical PRC, and is robust against noise. We confirmed this conclusion by applying different inference techniques to a high-dimensional system representing the simplest model of brain rhythm generation. Estimating the error, we have demonstrated that IPID-1 is the only approach to yield a good result for this challenging test.

We conclude the discussion of the phase response inference with a remark. While testing the algorithms on the model systems, we ensured that the stimulation was weak. In practice, one must perform stimulation with stimuli of different strengths and check whether the inferred PRC depends on this strength. No dependence means validity of the linear approximation and, hence, of the PRC description. Otherwise, the revealed curve is not the theoretical (infinitesimal) PRC but quantifies the effect of a finite-strength perturbation and is amplitude-dependent. 

\subsection{Amplitude response}
The main problem with the amplitude response's inference is the definition of the amplitude. While phase can be uniquely defined for any point in the basin of attraction of a limit cycle, the definition of the amplitude is ambiguous. An operational approach introduces the (empirical) amplitude variable as an envelope to the observable and computes the envelope employing, e.g., the Hilbert Transform. However, this definition obviously depends on the observable. Moreover, for non-harmonic signals, HT and other techniques generally do not provide a ``good'' envelope since they show variation for perfectly periodic signals.  

Another approach involves equating the amplitude with the slowest decaying isostable variable, which provides a universal description independent of the choice of observable.
The IPID-1 approach developed here reconstructs the first-order isostable dynamics directly from a scalar time series. The advantage of this technique is that inferred Eq.~(\ref{eq:iso}) can be further exploited to predict the effect of stimuli of a different shape. Furthermore, like the PRC inference, the technique provides an error measure that can be computed solely from data. We also mention the limitations of the procedure. It seems that inferring the isostable response is generally harder than inferring the phase response, and this is reflected in the performance of the IPID-1 technique by, e.g., comparing the inference where the ground truth is known. We also stress that Eq.~(\ref{eq:iso}) suffices for describing the dynamics only if the oscillator is two-dimensional or generally high-dimensional, but perturbations' decay in one direction is much slower than in other directions. The reduction of limit-cycle oscillators to phase-isostable dynamics is rather recent~\cite{wilson2016} and the physical interpretation of the isostable structure is still a matter of discussion. Another problem is to relate the isostable variables to 
envelopes of observed signals. Generally,  inference and quantification
of the amplitude response remain a subject of further studies.

\subsection{Relevance for oscillatory dynamics control}
Knowledge of the phase and amplitude response allows efficient control of an oscillatory system. Indeed, probing the system by pulses of a known shape, we infer the Winfree Eq.~(\ref{eq:Winfree}) and then can exploit this equation to optimize the stimulus' shape, e.g.,  aiming to maximize the response. 

In many cases, the goal of the control is to suppress or enhance the oscillation. The most illustrative example is deep brain stimulation (DBS), aiming to quench the Parkinsonian tremor. The design of efficient control schemes requires the determination of vulnerable phases where the stimulation is most efficient. A possible approach relies on an adaptive control scheme~\cite{Montaseri_et_al-13,Rosenblum-20}. The alternative idea is to use first a test stimulation to infer the response properties and then exploit the corresponding curves to determine the proper stimulation phases~\cite{Duchet_et_al-20}. 
Maxima and minima of the IRC or ARC provide the optimal stimulation phases for correspondingly enhancing and suppressing the oscillation.
We emphasize that PRC alone does not yield the required information. Although a relationship between PRC and ARC has been demonstrated for some examples~\cite{Duchet_et_al-20}, 
generally, this relationship does not hold, see Appendix~\ref{app:relation_PRC_ARC} for details. 
Another idea is stimulating around zeros of $Z(\vp)$. If the phase response is zero, the stimulus acts along the isochron and modifies only the amplitude. This idea might be a good starting point, but it does not guarantee that this phase is optimal
- the final amplitude variation depends on isostables density.
We believe that the reconstruction of the isostable equation from data provides a means to design an optimal stimulation. Just as in the case of phase response, using the reconstructed Eq.~(\ref{eq:iso}) we obtain the amplitude responses to arbitrary stimuli and can exploit this equation to optimize the stimulus' shape. 

\begin{acknowledgments}
R.C. and E.T.K.M. acknowledge financial support from Deutsche Forschungsgemeinschaft (DFG, German Research Foundation), Project-IDs PI 220/21-1 and 424778381 – TRR 295, respectively. \end{acknowledgments}

\appendix

\section{Phase-isostable coordinates}
\label{app:phipsi}

Isostable coordinates extend the phase description of a limit cycle by adding variables $\psi_i$ that describe the decay or growth of deviations from the limit cycle in the isochronal hyperplane. In the vicinity of a limit cycle, any $N$-dimensional system can be represented in the isostable coordinate system
\begin{align}
    \dot{\vp} &= \omega\;, \\
    \dot{\psi_i} &= \kappa_i \psi_i\;, 
    \qquad i = 1, \dots, N-1\;,
    \label{eq:isostable_N_dim}
\end{align}
where $\w$ is the frequency, and $\kappa_i$ are Floquet exponents of the limit cycle.
For a system with a stable limit cycle, all Floquet exponents have a negative real part, $\Re(\kappa_i) < 0$. However, in many cases, it is sufficient to consider only one isostable variable $\psi_{\max}$ corresponding to the largest real part of its Floquet exponent. This two-dimensional approximation is justified if there is a clear separation of time scales and all neglected $\psi_i$ decay considerably faster than $\psi_{\max}$.
Restricting our consideration to only one isostable coordinate $\psi_{\max}$ complementing $\vp$, we also assume the corresponding  Floquet exponent to be real. It means that we do not address the case of the slowest mode described by a complex-conjugate pair of Floquet exponents. Thus $\kappa := \max_i \Re(\kappa_i)$ and below we omit the subscript ${\max}$ in the notation $\psi_{\max}$.  
For a detailed discussion, in particular, on how to interpret a general $N$-dimensional system of phase-isostable coordinates with pairs of complex conjugate Floquet exponents, see \cite{wilson2016, wilson2018, wilson2019}.

As one can check from the autonomous dynamical  Eqs.~\eqref{eq:isostable_N_dim}, all variables $\psi_i$ are invariant to scaling with some non-zero factor $c$ -- the dynamical equations remain unchanged after a change of variables $\Tilde{\psi}:= c\psi$. As soon as external perturbations $p(t)$ are introduced to the system via 
\begin{align}
    \dot{\vp} &= \w + G_\vp(\vp, \psi) p(t)\,, \\
    \dot{\psi} &= \kappa \psi + G_\psi(\vp, \psi) p(t) \,,
\end{align}
the scaling of isostable variable leads to the adjustment in the stimulation function
\begin{align}
    G_{\Tilde{\psi}}(\vp, \Tilde{\psi}) = c G_\psi(\vp, \frac{\Tilde{\psi}}{c}) \;.
\end{align}
Thus, the IRC (defined as the isostable amplitude response evaluated at the limit cycle) is also scaled as a result of the scaling of $\psi$:
\begin{align}
    I_{\Tilde{\psi}}(\vp) = G_{\Tilde{\psi}}(\vp, 0) = c I_{\psi}(\vp)\,.
\end{align}
The PRC is invariant under rescaling of the isostable $\psi$. In the following derivations, we keep the scaling factor $c$ to show where it appears in the equations.

\subsection{Isostable reduction for the SL system}
\label{app:phipsiSL}

The Stuart-Landau system, Eq.~\eqref{eq:SL_cartesian}, is the analytically solvable normal form of a Hopf bifurcation: For $\mu >0$, a radial limit cycle at $R_0=\sqrt{\mu}$ establishes with a basin of attraction being the entire phase space except for the fixed point at the origin. Thus, each point in the phase space, here given in polar coordinates $R$ and $\theta$, can be assigned a phase by the well-known formula
\begin{align}
    \vp(R,\theta) = \theta - \alpha \ln \left(\frac{R}{\sqrt{\mu}} \right)\,,
\end{align}
for which one can verify that $\dot{\vp} = \eta - \alpha \mu =: \omega$. Since the SL system is two-dimensional, one additional isostable $\psi$ suffices to describe the dynamics in the vicinity of the limit cycle completely. Hence, for the SL example, the isostable coordinate system is a one-to-one transformation for $\mathbb{R}\setminus \{0\}$. For the case of radial isochrons ($\alpha=0$), $\psi$ has already been derived in Ref.~\cite{wilson2018a} as 
\begin{align}
    \psi(R,\theta) = c(1 - \frac{\mu}{R^2}) \,.
\end{align}
Moreover, this coordinate transformation yields $\dot{\psi} = -2\mu \psi := \kappa \psi$ also in the case of non-radial isochrons ($\alpha \neq 0$). From this expression, one can check the system's dynamical properties: the roots of $\psi$ at $R=\sqrt{\mu}$ yield the coordinates of the limit cycle, and the divergence at $R=0$ indicates the border of the limit cycle's basin of attraction. Also, as $\psi$ converges to a finite value as $R \rightarrow \infty$, we deduce that it takes a finite time to reach the vicinity of the limit cycle for initial conditions with arbitrarily large $R$.

The SL system can be re-written using the dynamical parameters $\w$ and $\kappa$ instead of $\eta$ and $\mu$ as
\begin{equation}
\begin{aligned}
   \dot{x} &= -\w y - (x^2 + y^2 + \frac{\kappa}{2})(x-\alpha y) 
   +G_x(x,y)p(t)\;,\\
   \dot{y} &=  ~~\w x - (x^2 + y^2 + \frac{\kappa}{2})(y+\alpha x)
   +G_y(x,y)p(t)
   \label{eq:SL_cartesian_alternative} \;,
\end{aligned}
\end{equation}
where the functions $G_{x,y}(x,y)$ specify how the perturbation acts on the oscillator.

The knowledge of the analytical form of the coordinate transformation of $\psi$ and $\vp$ allows us to compute the PRC $Z(\vp)$ and IRC $I(\vp)$ also in a closed analytical form. Those are, by definition, the response curves of phase and isostable variables evaluated at the limit cycle, where $\psi = 0$. In general, the response curves are computed via
\begin{align}
    \begin{pmatrix}
        I(\vp) \\
        Z(\vp) \\
    \end{pmatrix}
    =
    \evalat[\Big]{
    \begin{pmatrix}
        G_\psi \\
        G_\vp \\
    \end{pmatrix}
    }{\psi = 0}
    = \evalat{J_{\text{isostable} \leftarrow \text{polar}}}{\psi = 0} \cdot \evalat{J_{\text{polar} \leftarrow \text{Cartesian}}}{\psi = 0} \cdot 
    \evalat{\begin{pmatrix}
        G_x \\
        G_y \\
    \end{pmatrix}}{\psi = 0}\;,
\end{align}
where $J_{\text{isostable} \leftarrow \text{polar}}$ is the Jacobian of the coordinate transformation from polar to isostable coordinates, and $J_{\text{polar} \leftarrow \text{Cartesian}}$ is the well-known Jacobian of the coordinate transformation from Cartesian to polar coordinates:
\begin{align}
    J_{\text{isostable} \leftarrow \text{polar}} 
    =  
    \begin{pmatrix}
    \partial_R \psi & \partial_\theta \psi \\
    \partial_R \vp & \partial_\theta \vp \\
    \end{pmatrix}
    =
    \begin{pmatrix}
    \frac{2\mu c}{R^3} & 0 \\
    -\frac{\alpha}{R} & 1 \\
    \end{pmatrix}
    \quad \;, \quad 
    J_{\text{polar} \leftarrow \text{Cartesian}}
    =
    \begin{pmatrix}
    \partial_x R & \partial_y R \\
    \partial_x \theta & \partial_y \theta \\
    \end{pmatrix}
    =  
    \begin{pmatrix}
    \cos(\theta) & \sin(\theta) \\
    -\frac{\sin(\theta)}{R} & \frac{\cos(\theta)}{R} \\
    \end{pmatrix}
    \,.
    \label{eq:Jacobians_coordinate_transforms}
\end{align}
Throughout this paper, we assume a state-independent stimulation, namely, $G_x=\cos(\beta)$ and $G_y=\sin(\beta)$. Hence we arrive at the following phase and isostable response curves for the SL system:
\begin{align}
    I(\vp) &= \frac{2c}{\sqrt{-\frac{\kappa}{2}}}\cos(\vp-\beta) \;,\\
    Z(\vp) &= -\frac{1}{\sqrt{-\frac{\kappa}{2}}} \left( \sin(\vp-\beta) + \alpha \cos(\vp-\beta)\right)\;.
\end{align}

For the rotationally invariant SL system, it is convenient to interpret the polar radius $R$ as the amplitude of the limit cycle oscillation. Thus, from the response curve of $R$, denoted as $G_R$, we infer the ARC as
\begin{align}
   A(\vp)= \frac{R_e}{R_s} 
    \approx \frac{R_s + fG_R(\vp)}{R_s} 
    = 1 + \frac{fG_R(\vp)}{R_s} \;,
\end{align}
approximating a short pulse with action $f$ by a Dirac delta function, hitting the system at phase $\vp$. For the SL system, the initial amplitude $R_s=\sqrt{\mu}$ is independent of $\vp$. For a stimulation that is state-independent in Cartesian coordinates, the response in $R$ is given by $G_R(\vp) = \cos(\vp-\beta)$ and thus the ARC simplifies to the expression
\begin{align}
A(\vp)\approx 1 + \frac{f \cos(\vp-\beta)}{\sqrt{\mu}}\;.
\end{align}

\subsection{Test model with a non-sinusoidal solution and known phase and isostable response}
\label{app:phipsiEM}

In order to construct more complex two-dimensional test models with known properties, we use the isostable coordinate system. Instead of trying to derive the phase and isostable variable from known systems, we give an explicit analytical expression of $\vp$ and $\psi$ in the first place. Since the roots of $\psi(\mathbf{x})$ are the location of the stable invariant set for $\kappa<0$, we can control the position and shape of the limit cycle. In particular, the coordinate transformation 
\begin{align}
    \psi(R,\theta) = c\Big(1 - \frac{q(\theta)}{R^2}\Big )
\end{align}
describes a limit cycle at $R_0=\sqrt{q(\theta)}$ for any strictly positive, $2\pi$-periodic function $q$ of $\theta$. Together with the phase defined as
\begin{align}
    \vp(R,\theta) = \theta - \alpha \ln \left(\frac{R}{\sqrt{q(\theta)}} \right) \;,
\end{align}
such that $\vp = \theta$ on the limit cycle, we obtain a generalized Stuart-Landau system. For $q(\theta)=-\frac{\kappa}{2}=\mu$ we recover the original SL system. Since the Jacobian $J_{\text{isostable} \leftarrow \text{polar}}$ can be computed in the same straightforward manner as for the SL system as
\begin{align}
    \evalat{J_{\text{isostable} \leftarrow \text{polar}}}{\psi=0}
    = 
     \evalat[\bigg]{
    \begin{pmatrix}
        \partial_R \psi & \partial_\theta \psi \\
        \partial_R \vp & \partial_\theta \vp \\
    \end{pmatrix}
    }{\psi=0}
    =
    \evalat[\bigg]{
    \begin{pmatrix}
        \frac{2cq(\theta)}{R^3} & -\frac{cq'(\theta)}{R^2} \\
        -\frac{\alpha}{R} & 1 + \frac{\alpha q'(\theta)}{2q(\theta)} \\
    \end{pmatrix}
    }{\psi=0}
    =
    \begin{pmatrix}
        \frac{2c}{\sqrt{q(\theta)}} & -\frac{cq'(\theta)}{q(\theta)} \\
        -\frac{\alpha}{\sqrt{q(\theta)}} & 1 + \frac{\alpha q'(\theta)}{2q(\theta)} \\
    \end{pmatrix}
    \quad 
    \;,
\end{align}
we can explicitly write the dynamical equations, both in polar
\begin{align}
    \dot{R} &= \frac{R}{2q(\theta)} \left( \kappa \left(1+\frac{\alpha q'(\theta)}{2q(\theta)} \right) \left(R^2-q(\theta) \right) + \w_0 q'(\theta) \right)\;,\\
    \dot{\theta} &= \w_0 + \frac{\alpha \kappa}{2q(\theta)}(R^2-q(\theta))
    \;,
\end{align}
and in Cartesian coordinates (where $q$ and $q'$ have to be read as functions of $\theta(x,y)$)
\begin{align}
    \dot{x} &= 
    \omega \left( \frac{xq'}{2q}-y \right)
    + \frac{\kappa}{2}\left(\frac{x^2+y^2}{q}-1\right)\left(x + \alpha \left( \frac{xq'}{2q}-y \right) \right)\;, \\
    \dot{y} &=  
    \omega \left( \frac{yq'}{2q}+x \right)
    + \frac{\kappa}{2}\left(\frac{x^2+y^2}{q}-1\right)\left(y + \alpha \left( \frac{yq'}{2q}+x \right) \right) \;.
\end{align}
Having the explicit Jacobians for the coordinate transformation, we obtain the PRC and IRC generally given by the response curves in polar angle $G_\theta$ and polar radius $G_R$ as 
\begin{align}
    I(\vp) &= 2c \mathcal{I}(\vp) \;, \\
    Z(\vp) &= \evalat{G_\theta}{\psi=0}(\vp) - \alpha \mathcal{I}(\vp) \;,
    \label{eq:gSL_PRC_IRC}
\end{align}
with the function $\mathcal{I}$ defined as
\begin{align}
    \mathcal{I}(\vp) 
    &= q^{-\frac{1}{2}}(\vp) \evalat{G_R}{\psi=0}(\vp) -
    \frac{q'(\vp)}{2q(\vp)} \evalat{G_\theta}{\psi=0}(\vp) 
    \;.
\end{align}
The polar angle $\theta$ serves as a protophase in this model since, on the limit cycle, $\theta$ and $\vp$ coincide despite not being equal in general. Thus, on the limit cycle where $\psi=0$, we have $\theta=\vp$ and we obtain the shape of the limit cycle $R_0=\sqrt{q(\vp)}$ as a function of phase $\vp$.

Note the relation from Eqs.~\eqref{eq:gSL_PRC_IRC}, stating that the IRC is proportional to the difference of PRC and the response curve of the polar angle $\evalat{G_\theta}{\psi=0}$ for non-zero non-isochronicity parameter $\alpha$: 
\begin{align}
     Z(\vp) - \evalat{G_\theta}{\psi=0}(\vp) 
     &= -\frac{\alpha}{2c} I(\vp)
     \;.
\end{align}
We emphasize this relation between PRC and IRC since it exists without further assumptions made about $G_\theta$ and $G_R$, i.e., how external stimulation enters the system, but is exclusively due to the autonomous dynamics of the system. However, despite the generalizing extension this model has added to the simple SL model, it still holds strong assumptions that lead to this conclusion, e.g., that phase $\vp$ and polar angle $\theta$ are equal on the limit cycle.

Assuming that the external stimulation enters state independently in Cartesian coordinates, we compute the polar angle response function as 
\begin{align}
    \evalat{G_\theta}{\psi=0}(\vp) 
    = -\frac{\sin(\vp-\beta)}{\sqrt{q(\vp)}}
\end{align}
and $\mathcal{I}$ as
\begin{align}
    \mathcal{I}(\vp)
    &= \frac{2\cos(\vp-\beta)q(\vp) + \sin(\vp-\beta)q'(\vp)}{2q^{\frac{3}{2}}(\vp)} 
    \;. 
\end{align}
For the particular example in Sec.~\ref{sec:test}, we use $q(\theta)=r + 2\cos^2(\theta)$, where $r$ is a positive parameter. We denote this particular model as the modified Stuart-Landau (mSL) model since it does not contain the original SL model as a special case. Its polar angle response curve is given by
\begin{align}
    \evalat{G_\theta^\text{mSL}}{\psi=0}(\vp) 
    = -\frac{\sin(\vp-\beta)}{\sqrt{r+2\cos^2(\vp)}}
\end{align}
and the function $\mathcal{I}$ as
\begin{align}
     \mathcal{I}(\vp) = \frac{(r+1)\cos(\vp-\beta) + \cos(3\vp-\beta)}{(r+2\cos^2(\vp))^\frac{3}{2}}
     \;.
\end{align}
Having these two function, the PRC in Eq.~\eqref{eq:mSL_PRC} and IRC in Eq.~\eqref{eq:mSL_IRC} are derived using the relations given in Eq.~\eqref{eq:gSL_PRC_IRC} for $c=1$.

\subsection{Relationship of phase and amplitude response curves \label{app:relation_PRC_ARC}}

\begin{figure}
    \includegraphics[width=0.9\textwidth]{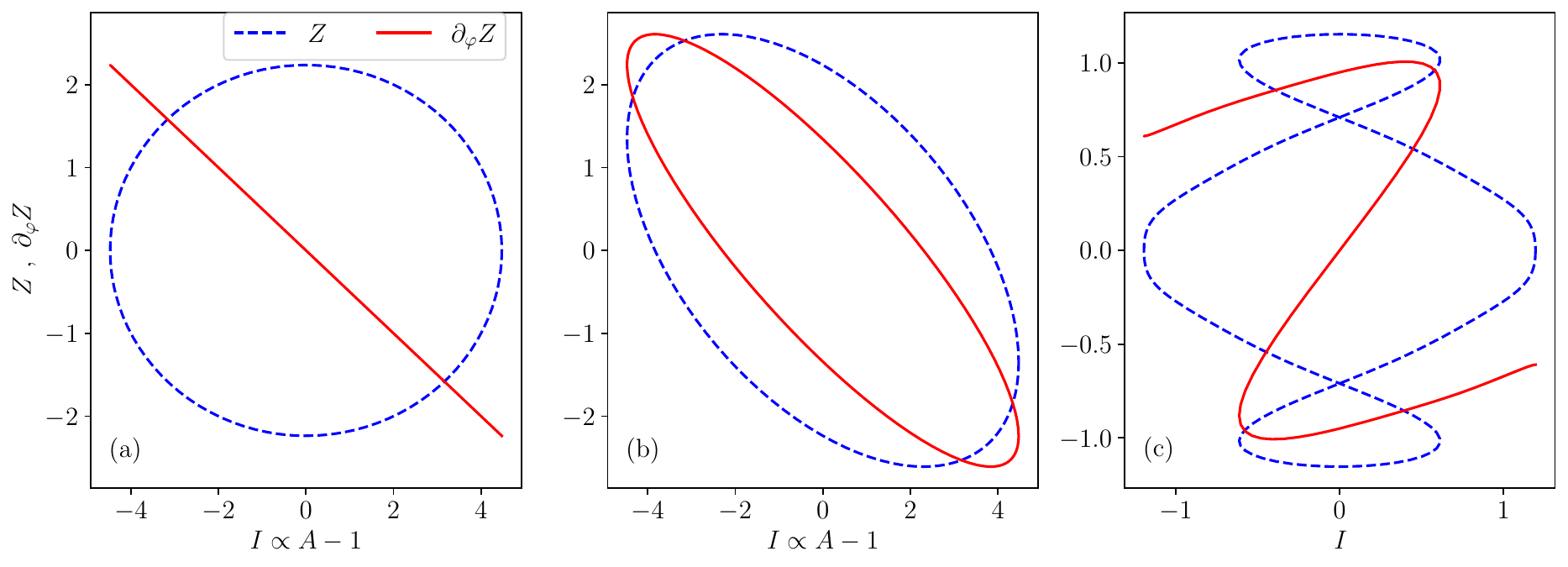}
    \caption{Relationships of phase and isostable response curves for the SL (Eqs.~\ref{eq:SL_PRC} and \ref{eq:SL_IRC}) and mSL (Eqs.~\ref{eq:mSL_PRC} and \ref{eq:mSL_IRC}) systems. In all panels, the blue dashed curve depicts the PRC, and the solid red curve is its derivative as a function of the IRC, parametrized by phase $\vp$. Panel (a): isochronous SL system with $\mu=0.2$ and $\alpha=0$. Panel (b): non-isochronous SL system with $\mu=0.2$ and $\alpha=0.6$. Panel (c): mSL system with $r=0.75$ and $\alpha=0$. For the SL oscillator, the amplitude can be expressed in terms of the polar radius $R$. Thus, in panel (a) and (b), the IRC is proportional to $A-1$ where $A$ is the ARC from Eq.~\eqref{eq:SL_ARC}. For the mSL depicted in panel (c), a straightforward definition of amplitude does not exist.}
    \label{fig:Response_curve_relations}
\end{figure}

Duchet et al.~\cite{Duchet_et_al-20} hypothesized that the amplitude response is proportional to $\partial_\vp{Z(\vp)}$. Indeed, Figure~\ref{fig:Response_curve_relations}a demonstrates that this hypothesis holds for the isochronous SL oscillator. 
In the following, we sketch how geometrical assumptions about an oscillating system can lead to this hypothesis on its own: Assume (as we also did throughout this paper) that in a two-dimensional system, the external stimulation is state-independent in Cartesian coordinates $x$ and $y$. Then, without loss of generality, the response functions can be written as $G_x = \rho \cos(\beta)$ and $G_y = \rho \sin(\beta)$, where the angle $\beta$ determines the direction of the stimulation in the $x$-$y$-plane and the non-negative value $\rho$ determines the strength. Without loss of generality, we can set $\rho=1$ by absorbing the stimulation strength as a factor into the stimulation shape function $p(t)$.
Using a coordinate transformation from Cartesian to polar coordinates, see Eq.~\eqref{eq:Jacobians_coordinate_transforms}, the response functions in the polar angle $\theta$ and the polar radius $R$ follow as 
\begin{align}
   G_\theta = -\frac{1}{R}\sin(\theta-\beta)
   \qquad \text{and} \qquad
   G_R = \cos(\theta-\beta) \;. 
   \label{eq:responses_polar}
\end{align} 
These two functions establish several relations, e.g.,
\begin{align}
    G_R = -R \cdot \partial_\theta G_\theta
    \qquad \text{or} \qquad
    R^2 G^2_\theta + G_R^2 = 1
    \,.
    \label{eq:relation_polar}
\end{align}
Thus, we obtain the relation that $G_R$ is proportional to the derivative of $G_\theta$. However, we emphasize that up to this point, the considerations are purely geometrical and are valid regardless of the system's dynamics. 

 If the limit cycle rotates around the origin, one sometimes tries to approximate the system's phase $\vp$ by the polar angle $\theta$ and the amplitude by the polar radius $R$. Indeed, for the standard SL system with its circular limit cycle, the polar radius $R$ can serve as a measure of amplitude, leaving the isostable response curve proportional to the response in $R$ and proportional to the ARC shifted by one: $I \propto G_R \propto A-1$. In addition, for the isochronous SL system ($\alpha = 0$, Fig.~\ref{fig:Response_curve_relations}a) the phase $\vp$ equals the polar angle $\theta$ and the geometric relation~\eqref{eq:relation_polar} directly translates into a relation between PRC and IRC: $I \propto \partial_\vp Z$.

For the non-isochronous SL system illustrated in Fig.~~\ref{fig:Response_curve_relations}b, this relation is no longer strict since phase $\vp$ and polar angle $\theta$ differ outside the limit cycle. Consequently, $Z \neq G_\theta$ and $\partial_\vp Z$ is only correlated instead of strictly proportional to $I$. The modified SL system (Fig.~\ref{fig:Response_curve_relations}c) has no circular limit cycle and hence no straightforward representation of amplitude by $R$, thus providing a counter-example to the hypothesis. 

\section{Relationship between empirical and infinitesimal PRC}
\label{app:inverse}
Here, we discuss how to recompute the empirical PRC 
$Z_{\cal P}$ into theoretical infinitesimal PRC $Z$.
The phase shift evoked by a pulse of finite width $\delta$ applied at the instant $t_s$ is 
$\Delta\vp=\vp(t_e)-\vp(t_s)-\w \delta$, where $t_e=t_s+\delta$. Hence, the Winfree Eq.~(\ref{eq:Winfree}) yields
$\Delta\vp=\int_{t_s}^{t_e}Z(\vp(t)){\cal P}(t)\dd t$.
If forcing is weak: $Z(\vp)p(t) \ll \omega$, then in the first approximation we take $\vp \approx \omega (t-t_s)$ and obtain:
\begin{equation}
    Z_{\cal P}\Big(\vp(t_s)\Big) = \frac{1}{f} \int_{t_s}^{t_e} Z\Big(\vp(t_s)+\omega (t-t_s)\Big) {\cal P}(t) \dd t\;.
    \label{eq:app2}
\end{equation}
where $f$ is the action of the pulse, in this paper either $f = \int_0^\delta {\cal P} \dd t $ for unipolar pulses, or $f = \frac{1}{2}\int_0^\delta |{\cal P}|\dd t $ for charge balanced pulses. 
This relation is a convolution of the true PRC $Z$ and the pulse's waveform ${\cal P}$: $Z_{\cal P} = Z \star {\cal P}$. 
The convolution operation can be expressed with a product relation in Fourier space: ${\cal F}(Z_{\cal P})={\cal F}(Z)\cdot {\cal F}({\cal P})$, which we utilize for expressing $Z(\vp)$:
\begin{equation}
    Z(\vp) = {\cal F}^{-1}\Big({\cal F}(Z_{\cal P})/{\cal F}({\cal P})\Big)\;,
\end{equation}
where ${\cal F}$ and ${\cal F}^{-1}$ denote the direct and inverse Fourier Transform, respectively.

We mention that for a charge-balanced pulse we do not recover the constant term in $Z$. Indeed, substituting in Eq.~(\ref{eq:app2}) $Z$ by $Z+C$, where $C$ is an arbitrary  constant, we obtain the same function $Z_{\cal P}$ due to 
the charge balance condition $\int_{t_s}^{t_e}{\cal P}\dd t=0$. To solve this problem, we set the mean value of $Z$ equal to that of $Z_{\cal P}$.

An alternative solution for the inverse problem that does not assume $\vp \approx \omega (t-t_s)$ is as follows. We represent the yet unknown function $Z(\vp)$ as a finite Fourier series with coefficients $z_n^{\cos}, z_n^{\sin}$. 
Then using $\mathcal{P}(t)$ we express $Z_{\mathcal{P}}(\vp)$ in terms of $z_n^{\cos}, z_n^{\sin}$ by numerically solving the Winfree equation. 
The nonlinear problem of equating the empirical curve with its symbolic representation: $Z_{\cal P}=Z_{\cal P}(\vp; z_n^{\cos}, z_n^{\sin})$, is then solved using the Levenberg-Marquardt technique~\cite{}.

\section{Thresholding a signal}
\label{app:threshold_crossing}

Here, we discuss how to determine events corresponding to the same asymptotic phase from the observed signal $s(t)$. 
This task is easy if the system moves on the limit cycle; then the signal $s(t)$ is a projection of a one-dimensional trajectory, and any defining event that occurs once per period will correspond to the same state and, therefore, the same phase. Generally, defining events can be threshold crossings, extrema, and inflection points. 

If the system is not on the limit cycle, determining the asymptotic phase is challenging. Generally, one would have to know the isochronal structure and then determine events as crossings of a particular isochron. (We remind that the isochrons are the manifolds of the constant phase.) However, even with {\it a priori} knowledge of isochrons, one cannot determine the crossing of a potentially high-dimensional manifold while observing a scalar signal $s(t)$ only. Therefore, we restrict our consideration to 2D oscillators or higher dimensional oscillators with one Floquet exponent much smaller in absolute value than the rest (having a slow 2D manifold). 
In general, for higher dimensional systems the approximation will be less accurate. Additionally, we mention that one does not need to find the events exactly, but just their good approximation. 

\begin{figure}
    \includegraphics[width=0.98\textwidth]{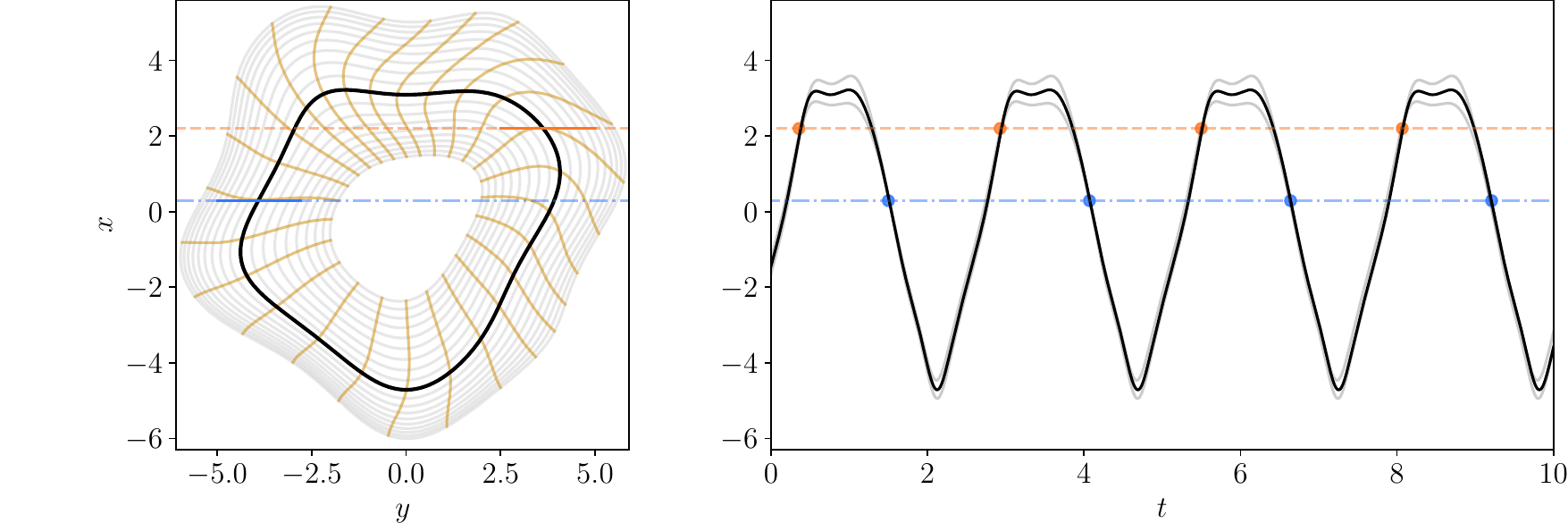}
    \caption{An illustration of choosing a proper threshold. (a) The isostable structure of a 2D limit-cycle oscillator. The limit cycle is depicted in black, and the isochrons are shown in yellow. The two thresholds tangential to isochrons are shown in red and blue. (b) The corresponding signal $x(t)$  is depicted in black. The two signals in gray correspond to a motion along a non-zero isochron. Note that the signal and two non-zero isochron signals all coincide at the ideal threshold crossings. Panels (a,b) share the vertical axis.}
    \label{fig:thresholding}
\end{figure}

Let us consider threshold crossings such as, e.g., $s(t) = s_\text{thr}, \frac{d}{dt}s(t) > 0$. Fig.~\ref{fig:thresholding} illustrates an example. Since we are thresholding a scalar signal $s(t)$ we only have one parameter: the threshold value $s_\text{thr}$. Different thresholds correspond to different parallel straight lines (see two examples in Fig.~\ref{fig:thresholding} depicted in red and blue). The best approximation in this situation is when the line is tangential to a particular isochron at the limit cycle. Then, the corresponding threshold provides a first-order approximation for the states with the same phase. While we cannot change the inclination of the line, we can choose its vertical position. For 2D systems, there always exist at least two such values, one when the signal is growing (see red dashed line in Fig.~\ref{fig:thresholding}) and one when it is falling (blue dash-dotted). From this illustration, one can deduce that extremal points are generally poor approximations for equal-phase states since the corresponding lines are (close to) tangential to the limit cycle and thus unlikely to be tangential to any isochron. 

\section{Test perturbation used in our experiments}
\label{app:forcing_gen}

Here, we discuss the test signals we apply to the test oscillators in our experiments. Except for the cases where we mention it explicitly, we consistently use a bipolar charge-balanced pulse with a period of positive forcing lasting 0.2, no forcing lasting 0.4, and a period of negative forcing lasting 1.0. The overall duration of the pulse is therefore 1.6, and since both our test oscillators~\eqref{eq:SL_cartesian} and \eqref{eq:infosc} are chosen to have a period 2$\pi$ each pulse lasts approximately $1/4$ of a period, see Fig.~\ref{fig:prc_standard} for a depiction of the pulse relative to the oscillators phase response curves. 
We use different forcing action $f=\frac{1}{2}\int_{0}^\delta |\mathcal{P}|dt$ for different oscillators: for the Stuart-Landau system~\eqref{eq:SL_cartesian} we used $f = 0.01$ while for the generalization~\eqref{eq:infosc} we used $f = 0.07$ since it has a larger limit cycle and thus needs a stronger perturbation for roughly the same effect. 

For the test with the standard inference technique, we introduced the pulses periodically with a much larger period (roughly five times) than the period of oscillators 2$\pi$, ensuring that the two periods are incommensurate; hence the oscillator would eventually be stimulated at almost every phase. Such perturbation is ideal for inferring the phase response since the pulses appear rarely, giving the oscillator time to relax to the limit cycle where the phase shift is easy to read. 
In contrast, for our introduced IPID-1 method~\ref{sec:phaseiso} we devised a more realistic experimental example where the pulses come randomly and often. We sample the inter-pulse times from a Poisson distribution such that, on average, one oscillator period receives 1.6 stimulating pulses, see Fig~\ref{fig:inference_example}b,h for a short signal vs. forcing example.

\section{Causal phase and amplitude estimation exploiting virtual oscillators}
\label{app:linosc}

Here, we show how to estimate phase and amplitude through a virtual oscillator. We explain how to obtain the amplitude first. 
We choose the oscillator's frequency $\eta$ to be much larger than the characteristic frequency $\nu$ of $s(t)$ and the damping parameter $\alpha_a$ sufficiently large so that the amplitude response
is practically independent of $\nu$.
Then, we recompute the amplitude of the virtual oscillator (that we can monitor) into the amplitude of the input signal, using the well-known expression for the resonance curve. Namely, we obtain the input's amplitude as
\[
a(t)=\sqrt{x^2+(\dot x/\nu)^2}\sqrt{(\eta^2-\nu^2)^2+(\alpha_a\nu)^2}\;.
\]

For phase estimation, we choose the damping parameter $\alpha_\vp$ to be small so that the phase shift $\beta=\arctan[-\alpha_\vp\nu/(\eta^2-\nu^2)]$ between the virtual oscillator and its input is nearly zero and weakly dependent on $\nu$ for $\nu\ll\eta$. Then, we compute input's phase as $\vp(t)=\arctan(-\dot x/\nu x)-\beta$.
For further detail and an efficient numerical scheme, we refer to Ref.~\cite{Rosenblum-Pikovsky-Kuehn-Busch-21}.

\section{Isostable inference step-by-step summary}
\label{sec:step_by_step}
\begin{enumerate}
    \item Determine events $\tau_i$ by thresholding the phase, e.g., $\vp(\tau_i) = \vp_\text{thr}$, $\frac{d}{dt}s(\tau_i) > 0$. The choice of threshold $\vp_\text{thr}$ is not crucial but generally should be picked such that the signal in events $s(\tau_i)$, has a wide range of values. 
    \item Write system~\eqref{eq:iso_int} by linearly approximating the isostable integral $\int_{\tau_i}^{\tau_{i+1}} (\psi(t)+s_0)dt \approx [s(\tau_i)+s(\tau_{i+1})](\tau_{i+1}-\tau_i)/2 $: 
    $$s(\tau_{i+1})-s(\tau_i) = -\kappa s_0(\tau_{i+1}-\tau_i) + \kappa \frac{s(\tau_i)+s(\tau_{i+1})}{2}(\tau_{i+1}-\tau_i) + \int\limits_{\tau_i}^{\tau_{i+1}} I(\vp(t)) p(t) \dd t\;.$$
    \item Express the response function $I(\varphi)$ as a Fourier series: $I(\vp) = \sum_{n=0}^{N_F} [u_n^{\cos} \cos(n\vp)+ u_n^{\sin} \sin(n\vp)]$, and replace the order of integration and summation in the last term to express it as a series of computable integrals:
    $$ \int\limits_{\tau_i}^{\tau_{i+1}} I(\vp(t)) p(t) \dd t =  \sum\limits_{n=0}^{N_F}[ u_n^{\cos} P_{n,i}^{\cos} + u_n^{\sin} P_{n,i}^{\sin}]\;,$$
    where we denoted the Fourier integrals with $P_{n,i}^{\cos} = \int_{\tau_i}^{\tau_{i+1}} \cos(n \vp(t)) p(t) \dd t$ and $P_{n,i}^{\sin} = \int_{\tau_i}^{\tau_{i+1}} \sin(n \vp(t)) p(t) \dd t$. Numerically compute the integrals for each inter-event interval.  
    \item Combine the approximations to express system~\eqref{eq:iso_int} as a linear system with known coefficients: 
    $$s(\tau_{i+1})-s(\tau_i) = -\kappa s_0(\tau_{i+1}-\tau_i) + \kappa \frac{s(\tau_i)+s(\tau_{i+1})}{2}(\tau_{i+1}-\tau_i) + \sum_{n=0}^{N_F}[  u_{n}^{\cos} P_{n,i}^{\cos} + u_{n}^{\sin} P_{n,i}^{\sin}]\;. $$
    There are as many equations as there are inter-event intervals. The known coefficients are: $s(\tau_{i+1})-s(\tau_i)$, $\tau_{i+1}-\tau_i$, $(s(\tau_{i})+s(\tau_{i+1}))/2$, and the Fourier integrals $P_{n,i}$. The unknown quantities are: $\kappa s_0,\ \kappa$ and the Fourier modes $u_n^{\cos},u_n^{\sin}$, representing the response $I(\vp)$.  
    Thus proceed to minimize this system via least squares (or a similar method) to obtain the first approximated solution: $s_0^{(1)}, \kappa^{(1)}, I^{(1)}(\vp)$.
    \item Iterate the following: approximate the continuous amplitude $\psi(t)$ by integrating the dynamics according to Eq.~\eqref{eq:integrate_iso} using the current approximation $s_0^{(m)},\kappa^{(m)},I^{(m)}(\vp)$. Numerically compute the integral $S_i = \int_{\tau_i}^{\tau_{i+1}} (\psi(t)+s_0) \dd t$ with the approximated amplitude. Solve system~\eqref{eq:iso_int} with the better approximated integral: 
    $$s(\tau_{i+1})-s(\tau_i) = -\kappa s_0(\tau_{i+1}-\tau_i) + \kappa S_i + \sum_{n=0}^{N_F} [ u_{n}^{\cos} P_{n,i}^{\cos} + u_{n}^{\sin} P_{n,i}^{\sin}] $$
    to obtain the next approximate solution $s_0^{(m+1)},\kappa^{(m+1)},I^{(m+1)}(\vp)$. The deviation of the reconstructed amplitude from the true amplitude estimated at events $\tau_i$ can be used as an error measure. 
\end{enumerate}



%

\end{document}